\documentstyle[12pt,epsf]{article}
\newcommand{\gtap}{\;{\raise.3ex\hbox{$>$\kern-.75em\lower1ex\hbox{$\sim$}}}\;}
\newcommand{\ltap}{\;{\raise.3ex\hbox{$<$\kern-.75em\lower1ex\hbox{$\sim$}}}\;}

\begin{document}
\begin{titlepage}

\rightline{hep-ph/0106293}
\smallskip
\rightline{June 26, 2001}
\bigskip\bigskip
\begin{center}
{\Large \bf Associated production of Higgs \\[7pt] and single top
at hadron colliders} \\
\medskip
\bigskip\bigskip\bigskip
{{\bf F.~Maltoni}, {\bf K.~Paul}, {\bf T.~Stelzer}, and {\bf S.~Willenbrock}} \\
\medskip
Department of Physics \\
University of Illinois at Urbana-Champaign \\ 1110 West Green Street \\
Urbana, IL\ \ 61801 \\
\bigskip
\end{center}
\bigskip\bigskip\bigskip

\vspace{.5cm}

\begin{abstract}
We study the production of the Higgs boson in association with a
single top quark at hadron colliders. The cross sections for the
three production processes ($t$-channel, $s$-channel, and
$W$-associated) at both the Tevatron and the LHC are presented. We
investigate the possibility of detecting a signal for the largest
of these processes, the $t$-channel process at the LHC, via the
Higgs decay into $b\bar{b}$. The QCD backgrounds are large and
difficult to curb, hindering the extraction of the signal.
Extensions of our analysis to the production of supersymmetric
Higgs bosons are also addressed. The cross section is enhanced for
large values of $\tan\beta$, increasing the prospects for
extracting a signal.
\end{abstract}

\vfil

\end{titlepage}

\section{Introduction}

The discovery of the Higgs boson as the culprit for Electroweak Symmetry
Breaking (EWSB) is one of the most challenging goals of present and future
high-energy experiments. Within the Standard Model (SM), the mass of the
physical Higgs particle is basically unconstrained with an upper bound of $m_h
\ltap 600-800$ GeV~\cite{Luscher:1988gc}. However, present data from precision
measurements of electroweak quantities  favor a moderate mass (113 GeV $< m_h
\ltap 200-230$ GeV)~\cite{LEPEWWG}. In addition, the minimal supersymmetric
version of the SM (MSSM), which is one of its most popular extensions,
predicts a Higgs boson with an upper mass bound of about 130 GeV
\cite{Haber:1991aw,Okada:1991vk,Ellis:1991nz}. Thus the scenario  with an
intermediate-mass Higgs boson (113 GeV $< m_h \ltap 130$ GeV) is both
theoretically plausible and well supported by the data.

Detailed studies performed for both the Tevatron and the LHC (see, for example,
Refs.~\cite{Carena:2000yx} and~\cite{TDRs}, respectively) have shown that there
is no single production mechanism or decay channel which dominates the
phenomenology over the intermediate-mass range for the Higgs.  Associated
production of $Wh$ or $Zh$~\cite{Glashow:1978ab} and
$t\bar{t}h$~\cite{Ng:1984jm,Kunszt:1984ri}, with the subsequent decay $h \to
\gamma \gamma$~\cite{Kleiss:1991vc,Marciano:1991qq,Gunion:1991kg} and $h \to b
\bar{b}$~\cite{Gunion:1985rv,Stange:1994ya,Stange:1994bb,Dai:1993gm,Goldstein:2001bp},
are presently considered the most promising reactions to discover an
intermediate-mass Higgs at both the Tevatron and the LHC. In this case one of
the top quarks or the weak boson present in the final state can decay
leptonically, providing an efficient trigger. The major difficulties in
extracting a reliable signal from either of these two channels are the
combination of a small signal and the need for an accurate control of all the
background sources. In this respect, it would be useful to have other
processes that could raise the sensitivity in this range of masses.

\begin{figure}[!t]
\begin{minipage}{1\textwidth}
\begin{center}
\vspace*{-1cm} \hspace*{0cm} \epsfxsize=8cm \epsfbox{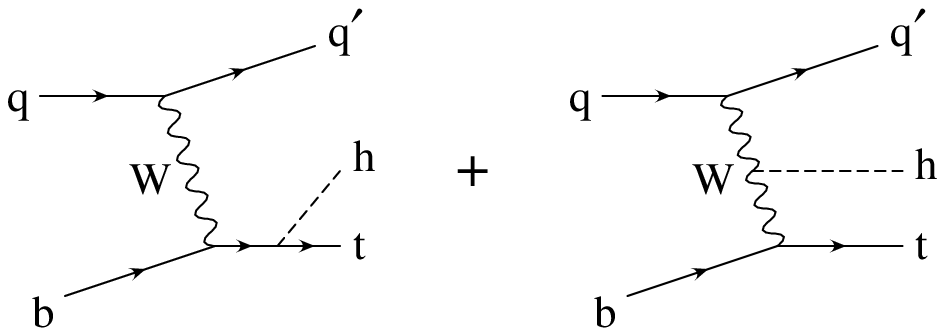} \vspace*{0cm}
\caption{Feynman diagrams contributing to the $t$-channel production of Higgs
plus single top.} \label{fig:t-channel}
\end{center}

%%%%%%%%%%%%%%%%%%%%%%%%%%%%%%%%%%%%%%%%%%%%%%%%%%%%%%%%%%%%%%%%%%%%%%%%%%%%%
%%%%%%%%%%%%%%%%%%%%%%%%%%%%%%%%%%%%%%%%%%%%%%%%%%%%%%%%%%%%%%%%%%%%%%%%%%%%%

\begin{center}
\vspace*{0cm} \hspace*{0cm} \epsfxsize=10cm \epsfbox{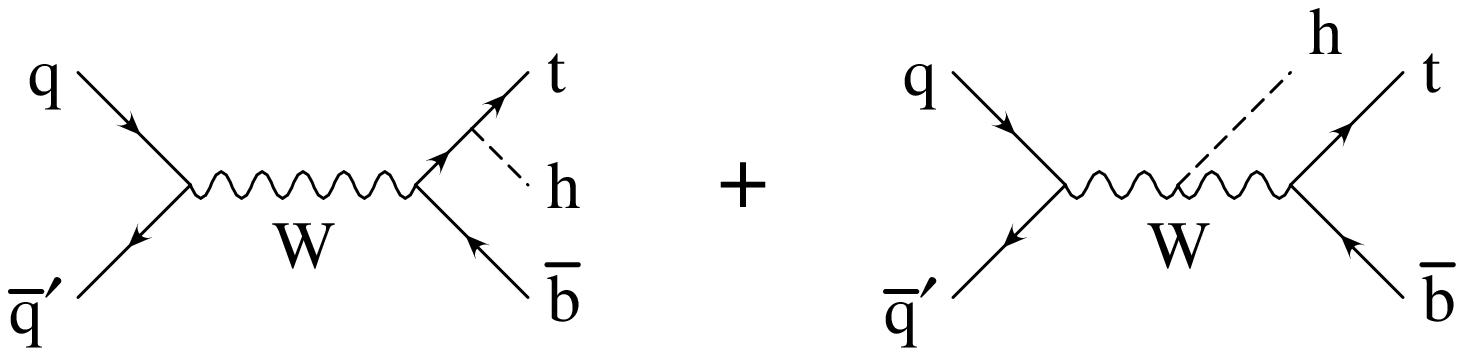} \vspace*{0cm}
\caption{Feynman diagrams contributing to the $s$-channel production of Higgs
plus single top.} \label{fig:s-channel}
\end{center}

%%%%%%%%%%%%%%%%%%%%%%%%%%%%%%%%%%%%%%%%%%%%%%%%%%%%%%%%%%%%%%%%%%%%%%%%%%%%%
%%%%%%%%%%%%%%%%%%%%%%%%%%%%%%%%%%%%%%%%%%%%%%%%%%%%%%%%%%%%%%%%%%%%%%%%%%%%%

\begin{center}
\vspace*{1cm} \hspace*{0cm} \epsfxsize=15cm \epsfbox{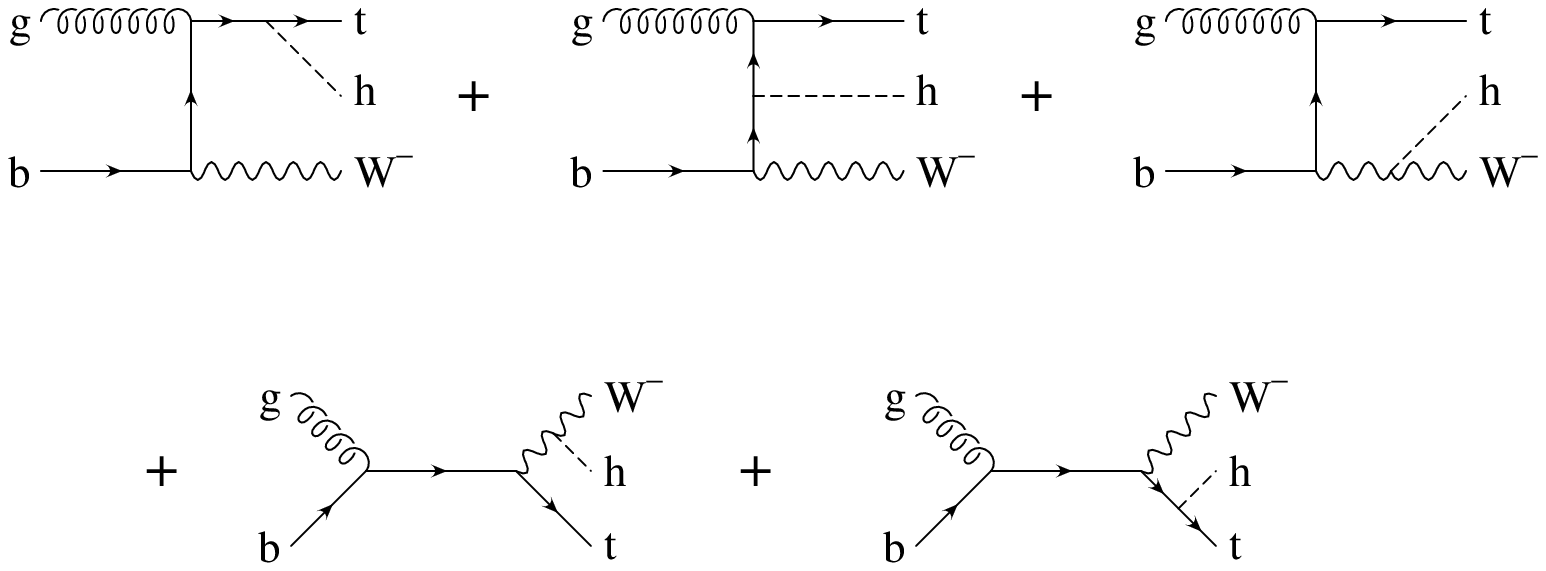}
\caption{Feynman diagrams contributing to the $W$-associated production of
Higgs plus single top.} \vspace*{0cm} \label{fig:W-associated}
\end{center}

%%%%%%%%%%%%%%%%%%%%%%%%%%%%%%%%%%%%%%%%%%%%%%%%%%%%%%%%%%%%%%%%%%%%%%%%%%%%%
\end{minipage}
\end{figure}

In this paper we revisit the production of a Higgs boson in association with a
single top quark ($th$ production) at hadron
colliders~\cite{Diaz-Cruz:1992cs,Stirling:1992fx,Ballestrero:1993bk,Bordes:1993jy}.\footnote{
We always understand $th$ to include both top and
anti-top production.} This process can be viewed as a natural extension of the
single-top production
processes~\cite{Willenbrock:1986cr,Yuan:1990tc,Ellis:1992yw,
Cortese:1991fw,Stelzer:1995mi,Heinson:1997zm}, where a Higgs boson is radiated
off the top or off the $W$ that mediates the bottom-to-top transition. As in
the usual single-top production, the three processes of interest are
characterized by the virtuality of the $W$ boson in the process: (i)
$t$-channel (Fig.~\ref{fig:t-channel}), where the spacelike $W$ strikes a $b$
quark in the proton sea, promoting it to a top quark; (ii)  $s$-channel
(Fig.~\ref{fig:s-channel}), where the $W$ is timelike; (iii) $W$-associated
(Fig.~\ref{fig:W-associated}), where there is emission of a real $W$ boson.

There are two reasons {\it a priori} that make the above processes worthy of
attention. The first one is that, based on simple considerations, one would
expect Higgs plus single-top production to be relevant at the Tevatron and at
the LHC. While top quarks will be mostly produced in pairs via the strong
interaction, the cross section for single top, which is a weak process, turns
out to be rather large, about one third of the cross section for top pair
production~\cite{Stelzer:1998ni,Beneke:2000hk}. If a similar ratio between
$\sigma(th)$ and $\sigma(t\bar{t}h)$ is assumed, it is natural to ask whether
$th$ production could be used together with $Wh$, $Zh$ and $t\bar{t}h$ as a
means to discover an intermediate-mass Higgs at the LHC. With this aim, the
$t$-channel process has been previously considered when the Higgs decays into a
pair of photons, with the result that too few events of this type would be
produced even at high-luminosity runs at the
LHC~\cite{Stirling:1992fx,Ballestrero:1993bk,Bordes:1993jy}. Since the dominant
decay mode of the Higgs in this mass region is into $b\bar{b}$ pairs, this
suggests searching for it using one or more $b$-tags, in a similar way as the
$t\bar{t}h$ analysis is conducted. This possibility is pursued in the present
paper.

%%%%%%%%%%%%%%%%%%%%%%%%%%%%%%%%%%%%%%%%%%%%%%%%%%%%%%%%%%%%%%%%%%%%%%%%%%%%%

\vfill
\pagebreak
\clearpage

The second reason for considering Higgs plus single-top production is that it
gives a rather unique possibility for studying the relative sign between the
coupling of the Higgs to fermions and to vector bosons
\cite{Bordes:1993jy,Tait:2001sh}. Measurements of $Wh$ and $t\bar{t}h$
production rates test respectively the Higgs coupling to the $W$ and the
Yukawa coupling to the top, but they cannot give any information on the
relative sign between the two. In the $th$ case, the $t$-channel and the
$W$-associated ($s$-channel) cross sections depend strongly on the destructive
(constructive) interference between the contributions from the Higgs radiated
off the top and off the $W$ boson. A measurement of the total rate for
production of Higgs plus single top would therefore provide additional
information on the EWSB sector of the SM.

As will be shown in detail in the following, at the Tevatron the cross section
for producing a Higgs in association with single top is of the order of $0.1$
fb and therefore out of the reach of Run II ($\ltap 15$ fb$^{-1}$). On the
other hand, with a cross section of the order of $100$ fb, several thousands of
events will be produced at the LHC with $30$ fb$^{-1}$. Whether this will be
enough to obtain a visible signal is the subject of the present investigation.
As we will see, the number of signal events left after branching ratios, cuts,
and efficiencies are taken into account is not large, and there are several
backgrounds, both irreducible and reducible, to consider.

This paper is organized as follows. In Section~\ref{sec:XSecs} we present the
leading-order results for Higgs plus single-top production at both the Tevatron
and the LHC, for the three channels mentioned above. The cross sections for
the $s$-channel and $W$-associated processes, as well as for the $t$-channel
process at the Tevatron, have not been presented before; we confirm the
$t$-channel cross section at the LHC calculated in
Refs.~\cite{Stirling:1992fx,Ballestrero:1993bk,Bordes:1993jy}. We investigate
in some detail the interference in the various channels.
Section~\ref{sec:tchannel} contains a study of signal and background for the
$t$-channel production at the LHC, with both three and four $b$-tags. Results
on the $t$-channel production at the LHC in the MSSM are discussed in
Section~\ref{sec:SUSY}. We present our conclusions in the last section.

\vskip1cm
\section{Cross Sections}
\label{sec:XSecs}

There are three channels for the production of Higgs plus single top at hadron
colliders:

\begin{center}
\begin{tabular}{llll}
 &$t$-channel     & \hspace*{1cm}$q b \to q' t h $       & \hspace*{1cm}(Fig.~\ref{fig:t-channel})\\
 &$s$-channel     & \hspace*{1cm}$q \bar{q}' \to \bar{b} t h $ & \hspace*{1cm}(Fig.~\ref{fig:s-channel})\\
 &$W$-associated  & \hspace*{1cm}$g b \to W^- t h $      & \hspace*{1cm}(Fig.~\ref{fig:W-associated})\\
\end{tabular}
\end{center}

\begin{figure}[t]
\begin{minipage}[t]{0.50\textwidth}
\begin{center}
\vspace*{0cm}
\hspace*{0cm}
\epsfxsize=8cm \epsfbox{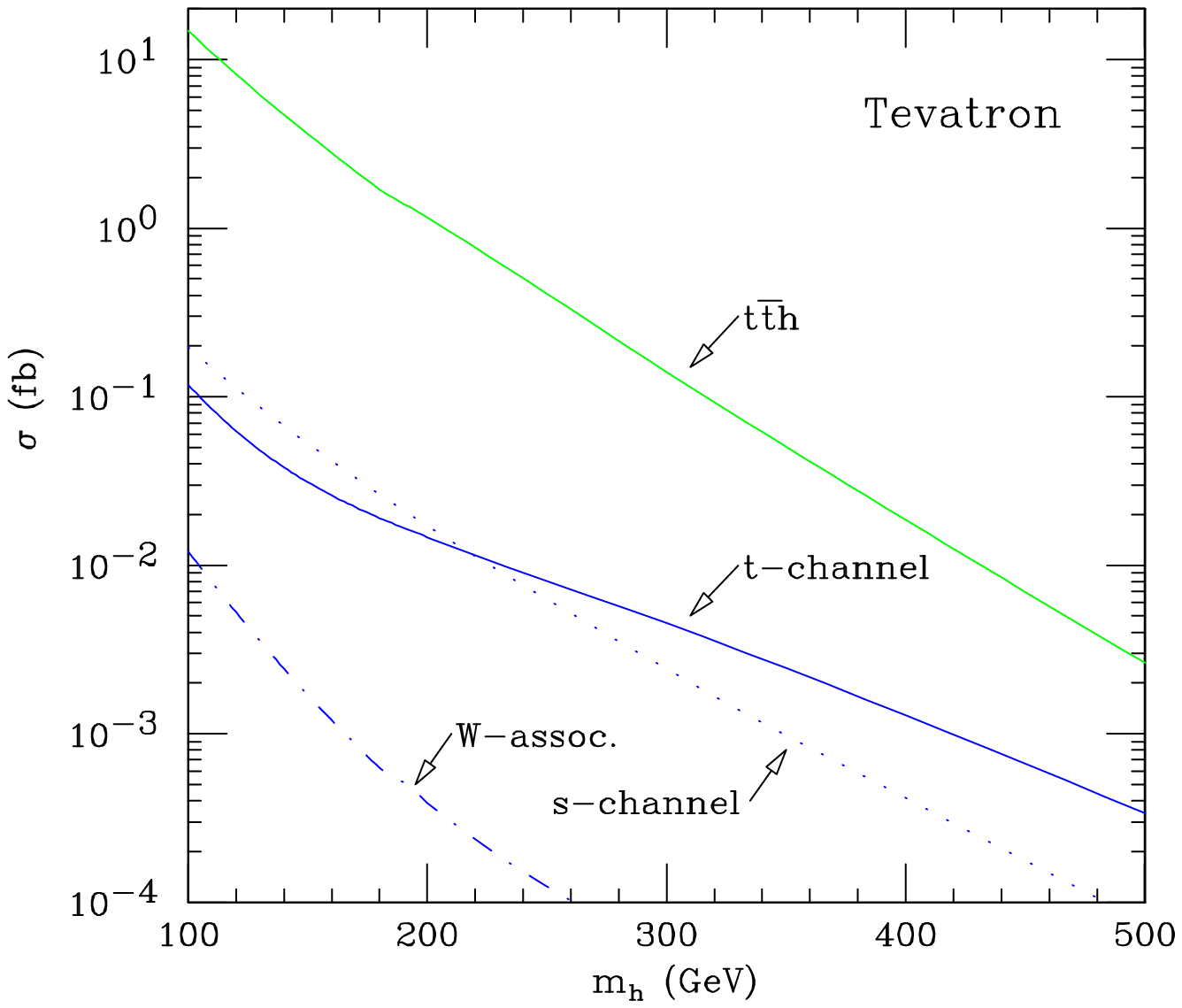}
\vspace*{0cm}
\end{center}
\end{minipage}
\hspace*{-0.3cm}
\begin{minipage}[t]{0.50\textwidth}
\begin{center}
\vspace*{0cm}
\hspace*{0cm}
\epsfxsize=8cm \epsfbox{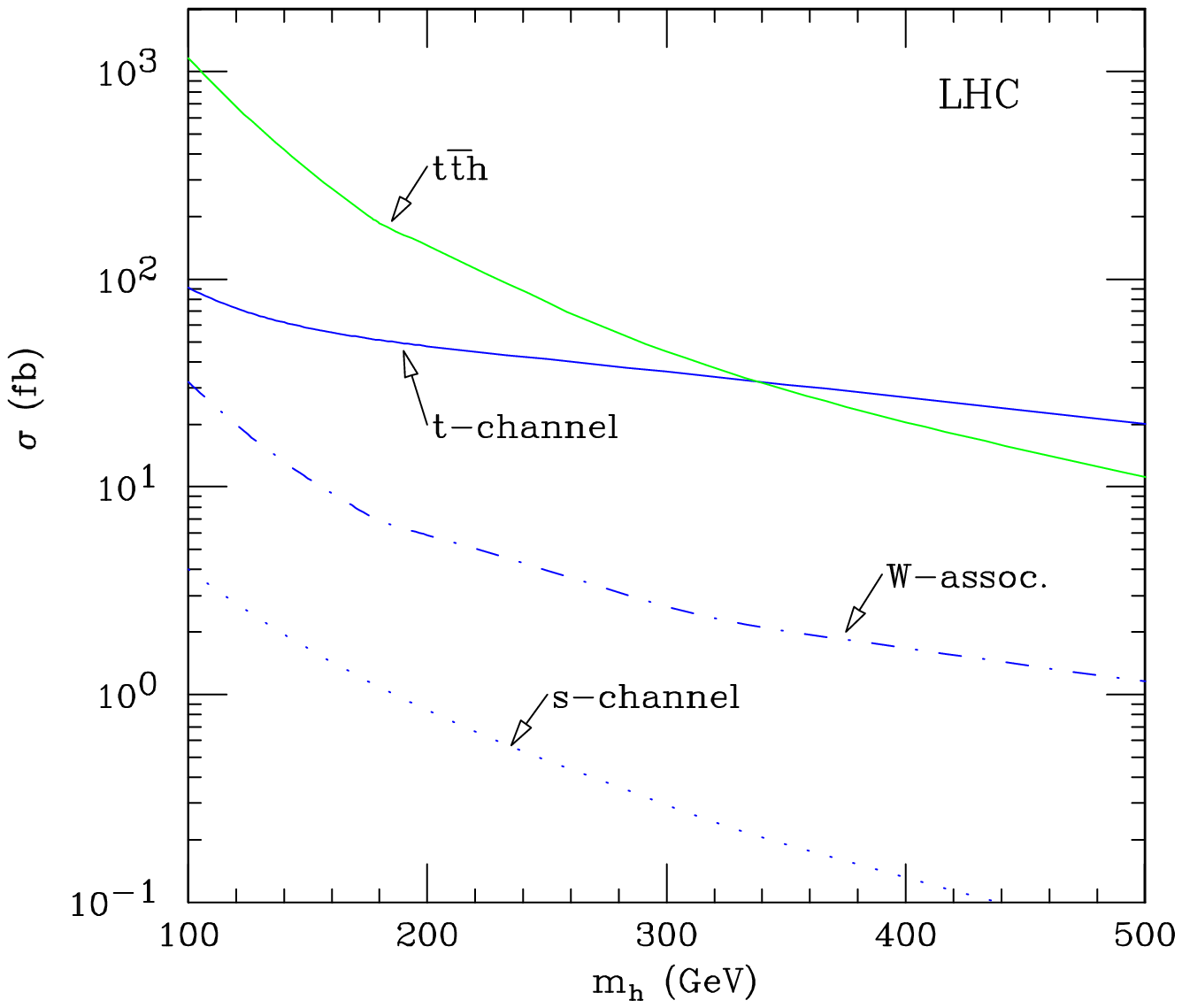}
\vspace*{0cm}
\end{center}
\end{minipage}
\caption{Cross sections for production of Higgs plus single top at the
Tevatron ($p\bar{p}$, $\sqrt{s}=2$ TeV) and at the LHC ($pp$, $\sqrt{s}=14$
TeV). Cross sections for the $t$-channel, $s$-channel
and $W$-associated  processes are shown. For comparison the cross
section for $t\bar{t}h$ is also shown. The set of parton
distribution functions is CTEQ5L, and the renormalization and factorization
scales are set equal to the Higgs mass.} \label{fig:xsec}
\end{figure}

\noindent In each case, the Higgs boson may be radiated off the top quark or
off the $W$ boson. Fig.~\ref{fig:xsec} shows the total cross section for each
channel at the Tevatron and at the LHC. These have been calculated using
tree-level matrix elements generated by MADGRAPH~\cite{Stelzer:1994ta} (and
checked against those obtained by COMPHEP~\cite{Pukhov:1999gg}) convoluted
with the parton distribution function set CTEQ5L~\cite{Lai:1995bb}, with the
renormalization\footnote{The renormalization scale is relevant only for the
$W$-associated process.} and factorization scales set equal to the Higgs
mass.\footnote{In the $t$-channel process, the factorization scale of the
light-quark distribution function should actually be the virtuality of the $W$
boson \cite{Stelzer:1997ns}. However, it happens that this makes little
difference numerically.}
%%%%%%%%%%%%%%%%%%%%%%%%%%%%%%%%%%%%%%%%%%%%%%%%%%%%%%%%%%%%%%%%%%%%%%%%%%%%%
%%%%%%%%%%%%%%%%%%%%%%%%%%%%%%%%%%%%%%%%%%%%%%%%%%%%%%%%%%%%%%%%%%%%%%%%%%%%%
%%%%%%%%%%%%%%%%%%%%%%%%%%%%%%%%%%%%%%%%%%%%%%%%%%%%%%%%%%%%%%%%%%%%%%%%%%%%%
\begin{table}[t]
\label{tab:comparison}
\caption{
Comparison of the ratios $\sigma(th)/\sigma(t)$ and
$\sigma(t\bar{t}h)/\sigma(t\bar{t})$, for a Higgs of mass $m_h=115\; {\rm
GeV}$, at the LHC and at the Tevatron. The set of parton distribution functions
is CTEQ5L, and the renormalization and factorization scales are set equal to
the top-quark mass in the $t$ and $t\bar{t}$ production and to the Higgs mass
in the associated processes. All results are leading order. In the second and
fourth line, ``$t$-Higgs only'' means that only the contribution where the
Higgs couples to the top (first diagram in Figs.~\ref{fig:t-channel}
and~\ref{fig:s-channel}) is included in the calculation of $\sigma(th)$.}
\medskip
\addtolength{\arraycolsep}{0.1cm}
\renewcommand{\arraystretch}{1.4}
\begin{center}
\begin{tabular}[4]{c|ccc|cc}
\hline\noalign{\vskip2pt}\hline
& & \multicolumn{1}{c}{$\sigma(th)/\sigma(t)\cdot 10^3$ }& &
  \multicolumn{2}{c}{$\sigma(t\bar{t}h)/\sigma(t\bar{t})\cdot 10^3$  }    \\
             & & $t$-ch     &   $s$-ch    & $ gg$ & $q\bar{q}$\\
\hline
         & & 0.33       &  0.42       & 1.1   & 3.1       \\[-4mm]
       LHC & & & &\\[-4mm]
       &$t$-Higgs only&1.1  & 0.28 &       &           \\
\hline
    & &0.038      & 0.20        & 0.26  & 1.6       \\[-4mm]
Tevatron & & & & \\[-4mm]
& $t$-Higgs only&0.21& 0.14 &       &           \\
 \hline  \hline
\end{tabular}

\end{center}
\end{table}
%%%%%%%%%%%%%%%%%%%%%%%%%%%%%%%%%%%%%%%%%%%%%%%%%%%%%%%%%%%%%%%%%%%%%%%%%%%%%%%%%%%%%%%
At the Tevatron, the $s$-channel process is enhanced
by the $p\bar{p}$ initial state and the relatively-low machine energy, and its
contribution is of the same order of magnitude as that of the $t$-channel
process. In contrast, the $t$-channel process dominates at the LHC. For the
sake of comparison, we have included in Fig.~\ref{fig:xsec} the rates for
production of a Higgs in association with a $t\bar{t}$ pair.

For intermediate-mass Higgs bosons, $\sigma(th)$ is much smaller than $\sigma
(t \bar{t} h)$, their ratio being $\sim 1/10$ at the LHC and $\sim 1/50$ at the
Tevatron. This is surprising since the analogous ratio between single-top and
$t\bar{t}$ production is $\sim 1/2$ at both the LHC and the
Tevatron.\footnote{As mentioned in the Introduction, the theoretical prediction
for the ratio $\sigma(t)/\sigma (t \bar{t})$ at the Tevatron and the LHC
is $\sim 1/3$, when calculated at next-to-leading order in the strong
coupling~\cite{Stelzer:1998ni,Beneke:2000hk}. However, since our results for
associated production of Higgs plus single top are only at tree-level, we
compare quantities evaluated at the lowest order.}

\begin{figure}[p]
\begin{minipage}[t]{0.50\textwidth}
\begin{center}
\vspace*{0cm}
\hspace*{0cm}
\epsfxsize=8cm \epsfbox{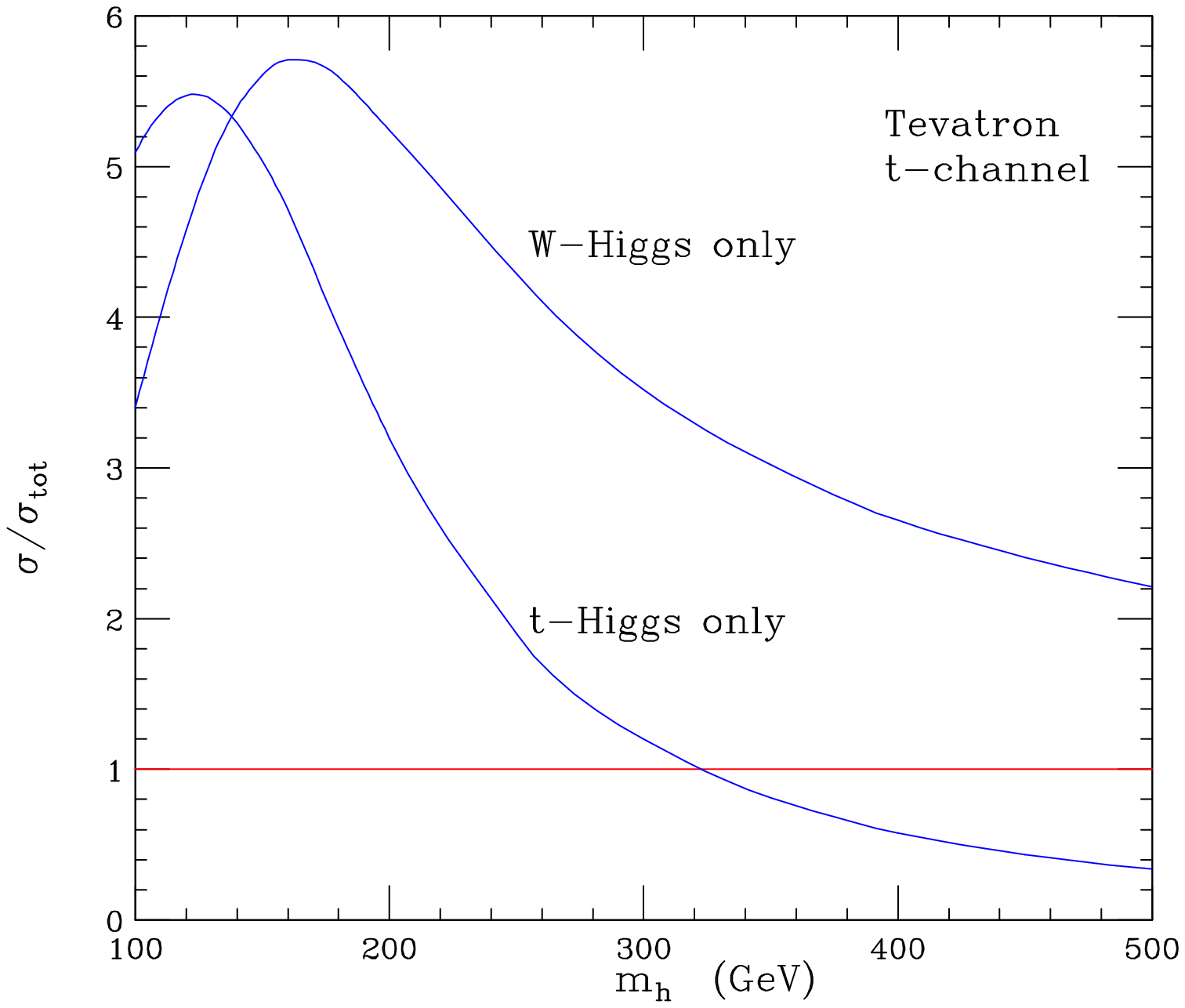}
\vspace*{0cm}
\end{center}
\end{minipage}
%%%%%%%%%%%%%%%%%%%%%%%%%%%%%%%%%%%%%%%%%%%%%%%%%%%%%%%%%%%%%%%%%%%%%%%%%%%%%
\hspace*{-0.5cm}
%%%%%%%%%%%%%%%%%%%%%%%%%%%%%%%%%%%%%%%%%%%%%%%%%%%%%%%%%%%%%%%%%%%%%%%%%%%%%
\begin{minipage}[t]{0.50\textwidth}
\begin{center}
\vspace*{0cm}
\hspace*{0cm}
\epsfxsize=8cm \epsfbox{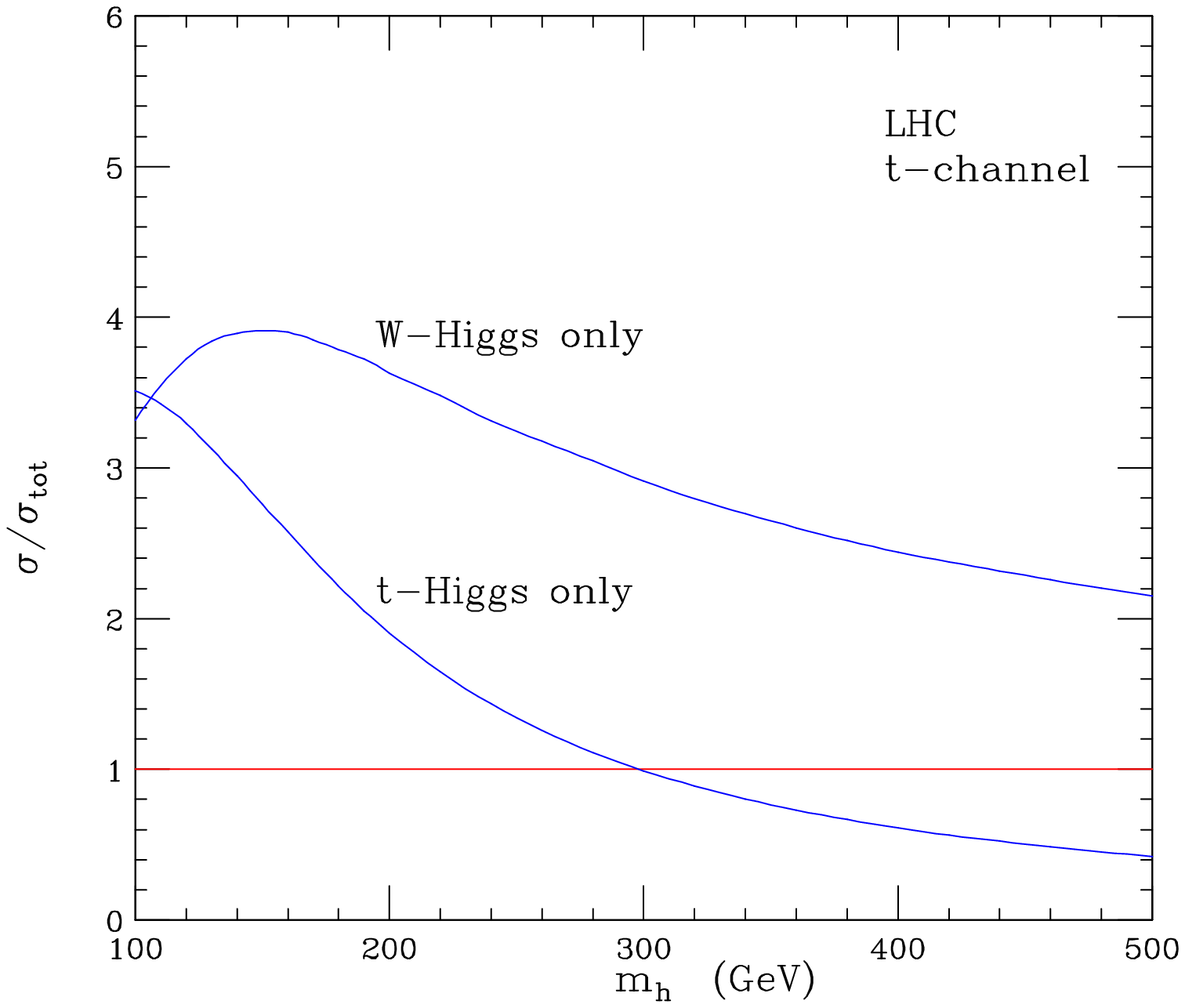}
\vspace*{0cm}
\end{center}
\end{minipage}
\caption{Interference in the $t$-channel process at the Tevatron and at the
LHC. The contributions from the $t$-Higgs coupling only and the $W$-Higgs
coupling only, normalized to the total cross section at any given Higgs mass,
are shown.} \label{fig:t-inter}
\end{figure}

It is instructive to pin down the reason for this strong suppression. With this
aim we compare in Table~1 the ratio of the cross sections
for single top, $\sigma(t)$, and for a $t\bar{t}$ pair, $\sigma(t \bar{t})$,
with the ratio where the Higgs is also produced, $\sigma(th)$ and $\sigma(t
\bar{t}h)$. We explicitly single out the contributions from different
channels, since their relative importance changes with the collision energy
and initial-state particles.  Looking at the leading contributions at the LHC
($t$-channel for single top and $gg\to t\bar t$) in the first line, we find a
suppression factor between the two processes of about $0.33/1.1\simeq 0.3$.
This is due to the destructive interference between the two diagrams in
Fig.~\ref{fig:t-channel}~\cite{Bordes:1993jy,Tait:2001sh}.\footnote{The
separation of the amplitude into contributions coming from the Higgs coupling
to the top quark and to the $W$ is gauge invariant. In the unitary gauge this
corresponds to considering the two diagrams in Fig.~\ref{fig:t-channel}
independently.} In Fig.~\ref{fig:t-inter} we have plotted the relative
contributions to the $t$-channel cross section from each of the two diagrams
in Fig.~\ref{fig:t-channel},
as a function of the Higgs mass, at the Tevatron and at the LHC. At the LHC,
for a Higgs mass of $115$ GeV, the cross section due to each diagram alone is
$\simeq 3.5$ times larger than the complete cross section, while, for larger
Higgs masses, the $W$-Higgs contribution becomes dominant.\footnote{ This
diagram contains a term proportional to the Higgs mass itself, as can be seen
by calculating the contribution coming from the exchange of a longitudinal $W$
in the $t$-channel. It is exactly this term that dominates the amplitude at
large Higgs masses.} To further support this argument, we have included the
contributions to $\sigma(th)$ coming from only the first diagram in
Fig.~\ref{fig:t-channel} in the second and fourth lines of Table~1. Comparing again the
ratio $\sigma(th)/\sigma(t)$ in the $t$-channel with the $gg$
contribution to $\sigma(t\bar{t}h)/\sigma(t \bar{t})$ at the LHC, we
find that they are the same $(1.1\cdot 10^{-3})$. Hence the suppression factor
of about $0.3$ found before is accounted for by the destructive interference.
The same argument applies at the Tevatron ($0.21 \cdot 10^{-3}\simeq 0.26\cdot
10^{-3}$), where the destructive interference is somewhat stronger than at the
LHC ($0.038/0.21 \simeq 0.18$) (Fig.~\ref{fig:t-inter}).

%%%%%%%%%%%%%%%%%%%%%%%%%%%%%%%%%%%%%%%%%%%%%%%%%%%%%%%%%%%%%%%%%%%%%%%%%%%%%
%%%%%%%%%%%%%%%%%%%%%%%%%%%%%%%%%%%%%%%%%%%%%%%%%%%%%%%%%%%%%%%%%%%%%%%%%%%%%
%%%%%%%%%%%%%%%%%%%%%%%%%%%%%%%%%%%%%%%%%%%%%%%%%%%%%%%%%%%%%%%%%%%%%%%%%%%%%
\begin{figure}[p]
\begin{minipage}[t]{0.50\textwidth}
\begin{center}
\vspace*{0cm}
\hspace*{0cm}
\epsfxsize=8cm \epsfbox{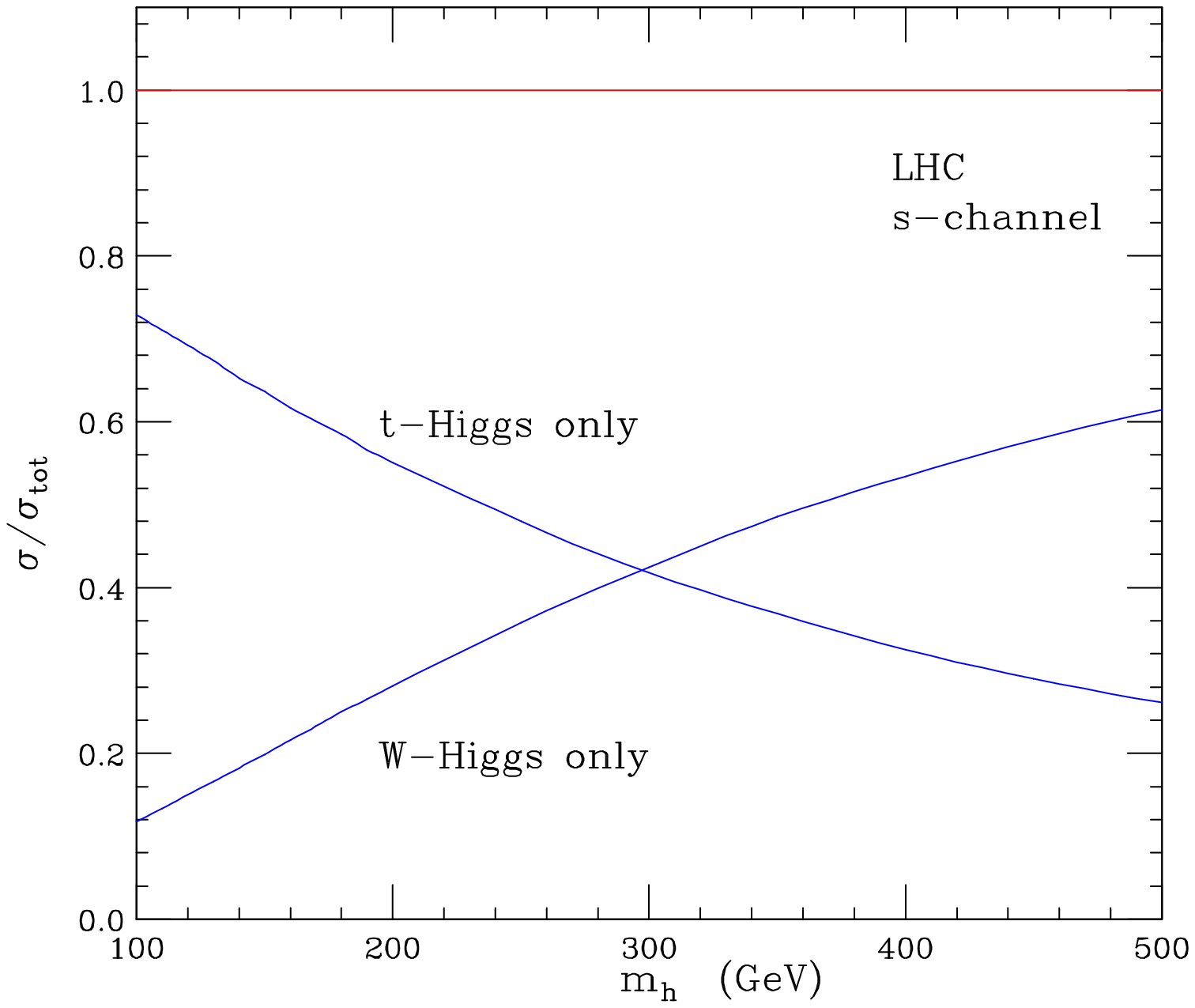}
\vspace*{0cm}
\end{center}
\end{minipage}
%%%%%%%%%%%%%%%%%%%%%%%%%%%%%%%%%%%%%%%%%%%%%%%%%%%%%%%%%%%%%%%%%%%%%%%%%%%%%
\hspace*{-0.5cm}
%%%%%%%%%%%%%%%%%%%%%%%%%%%%%%%%%%%%%%%%%%%%%%%%%%%%%%%%%%%%%%%%%%%%%%%%%%%%%
\begin{minipage}[t]{0.50\textwidth}
\begin{center}
\vspace*{0cm}
\hspace*{0cm}
\epsfxsize=8cm \epsfbox{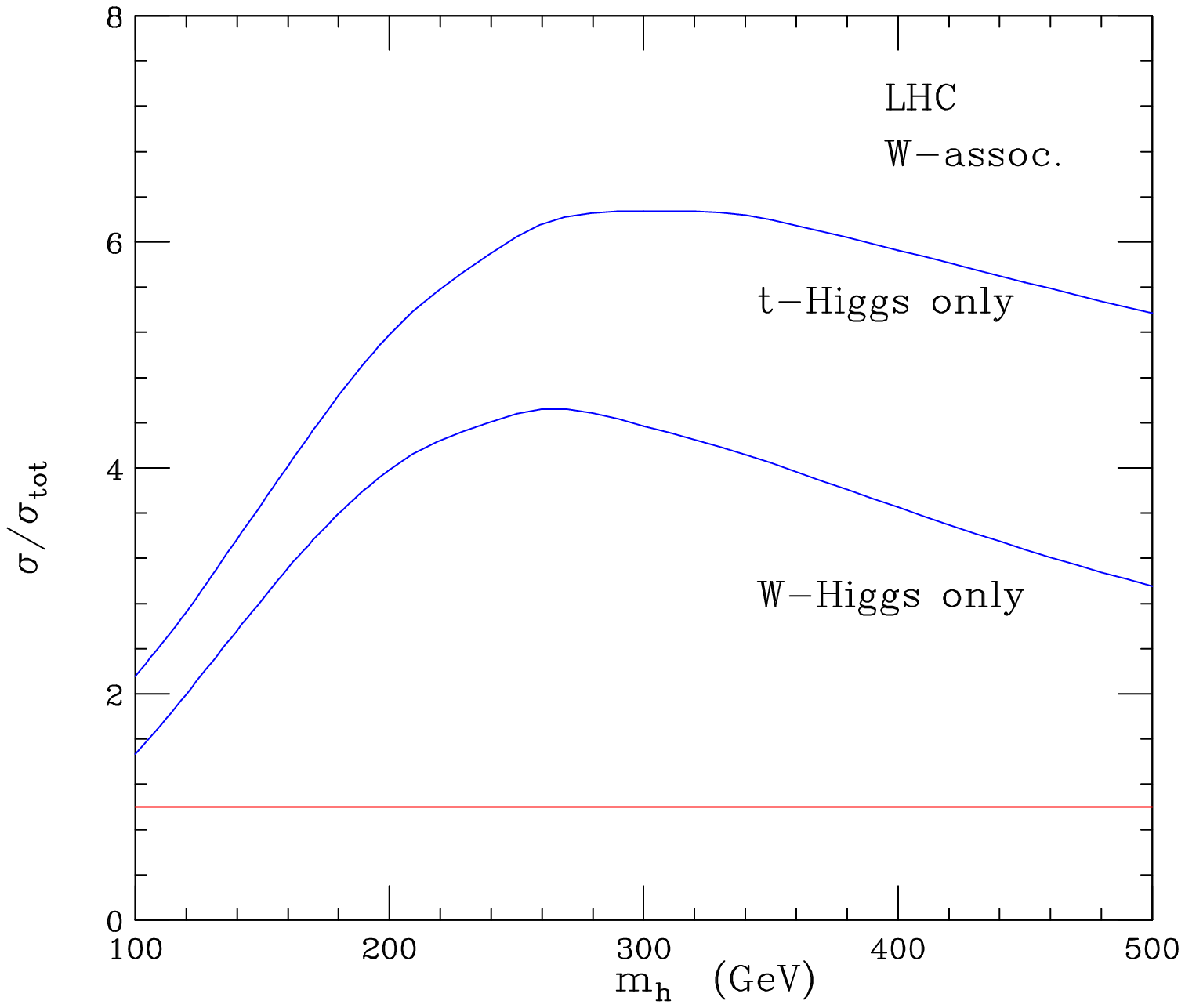}
\vspace*{0cm}
\end{center}
\end{minipage}
\caption{Interference in the $s$-channel (left) and in the $W$-associated
channel (right) at the LHC. The contributions from  the $t$-Higgs coupling
only or $W$-Higgs coupling only, normalized to the total cross section at any
given Higgs mass, are shown.} \label{fig:sW-inter-LHC}
\end{figure}
%%%%%%%%%%%%%%%%%%%%%%%%%%%%%%%%%%%%%%%%%%%%%%%%%%%%%%%%%%%%%%%%%%%%%%%%%%%%%

As can be seen from Fig.~\ref{fig:t-inter}, the reduction of the cross section
due to this interference effect strongly depends on the mass of the Higgs. In
this respect the large suppression found for Higgs masses less than $200$ GeV
can be regarded as a numerical accident. On the other hand, the fact that the
interference is destructive is a consequence of unitarity \cite{Bordes:1993jy}.
The simplest way to show this it to recall  that at high energies one can
describe the $t$-channel process in the so-called effective-$W$
approximation~\cite{Dawson:1985gx,Kunszt:1988tk}, where the initial light quark
emits a $W$ which may be treated as if it is on shell. In so doing the diagram
can be factorized into a distribution function of the $W$ in the initial quark
times a $2 \to 2 $ subprocess $W b \to h t $. One can show that at high
energies $E$, with $s \sim -t \sim -u \sim E^2 \gg m_h^2,m_W^2,m_t^2$, each of
the two sub-diagrams in Fig.~\ref{fig:t-channel} behaves like
\begin{equation}
{\cal A}_{\rm t-ch}^{t,W} \sim \, g^2\, \frac{
m_t E}{m_W^2}\,, \label{eq:t-unitarity}
\end{equation}
for an external longitudinal $W$, where the superscripts $t$ and $W$ indicate
from which particle the Higgs is radiated. For a $2 \to 2 $ process unitarity
demands that the total amplitude approaches at most a constant and therefore the
terms in Eq.~(\ref{eq:t-unitarity}) would violate unitarity at a scale
$\Lambda \simeq m_W^2/m_t g^2$. However, the unitarity-violating terms in the
two amplitudes have opposite signs and cancel when the two diagrams are added.
We conclude that although the amount of the suppression depends on the
parameters describing the process (such as the top mass, the Higgs mass, and
the center-of-mass energy), the sign of the interference term is a fundamental
property of the Higgs sector of the standard model. Moreover, we expect that
in extensions of the standard model where unitarity is respected up to
arbitrarily high scales, similar cancellations take place. As an example, we
have considered the $t$-channel production in a generic two-Higgs-doublet
model (2HDM) and explicitly verified that the terms that grow with energy
cancel. The details are presented in Appendix A.

There is a similar explanation of the cancellation between diagrams
in the $W$-associated production. At high energies the two gauge-invariant
classes of amplitudes, ${\cal A}^{t}$ and ${\cal A}^{W}$, behave like
\begin{equation}
{\cal A}_{\rm W-assoc.}^{t,W} \sim \, g_s \, g^2 \frac{m_t}{m_W^2}
\label{eq:W-unitarity}
\end{equation}
for an external longitudinal $W$.  Since for a $2 \to 3$ process unitarity
demands that the total amplitude decreases as $1/E$, a violation would occur at the
scale $m^2_W/m_t g^2 g_s$. We explicitly verified that the terms
in Eq.~(\ref{eq:W-unitarity}) cancel when the amplitudes are added together. In
the $s$-channel process, where the interference is constructive, the $W$ always
has a large timelike virtuality and the diagrams do not contain any divergent
behavior with energy.  The interference in the $s$-channel and $W$-associated
processes is demonstrated in Fig.~\ref{fig:sW-inter-LHC}.

\vskip1cm
\section{$t$-channel production at the LHC}
\label{sec:tchannel}

In this section we discuss  whether a signal for Higgs plus single top can be
disentangled from the backgrounds. As we have seen in the previous section,
the cross section at the Tevatron is far too small to be relevant and
therefore we do not investigate it any further. Here we focus on production at
the LHC and in particular on the $t$-channel process which is the dominant
contribution. All signal and background cross sections are calculated using
MADGRAPH~\cite{Stelzer:1994ta}.

\begin{figure}[tbp]
\begin{minipage}[t]{0.50\textwidth}
%%%%%%%%%%%%%%%%%%%%%%%%%%%%%%%%%%%%%%%%%%%%%%%%%%%%%%%%%%%%%%%%%%%%%%%%%%%%%
\begin{center}
\vspace*{0cm}
\hspace*{0cm}
\epsfxsize=8cm \epsfbox{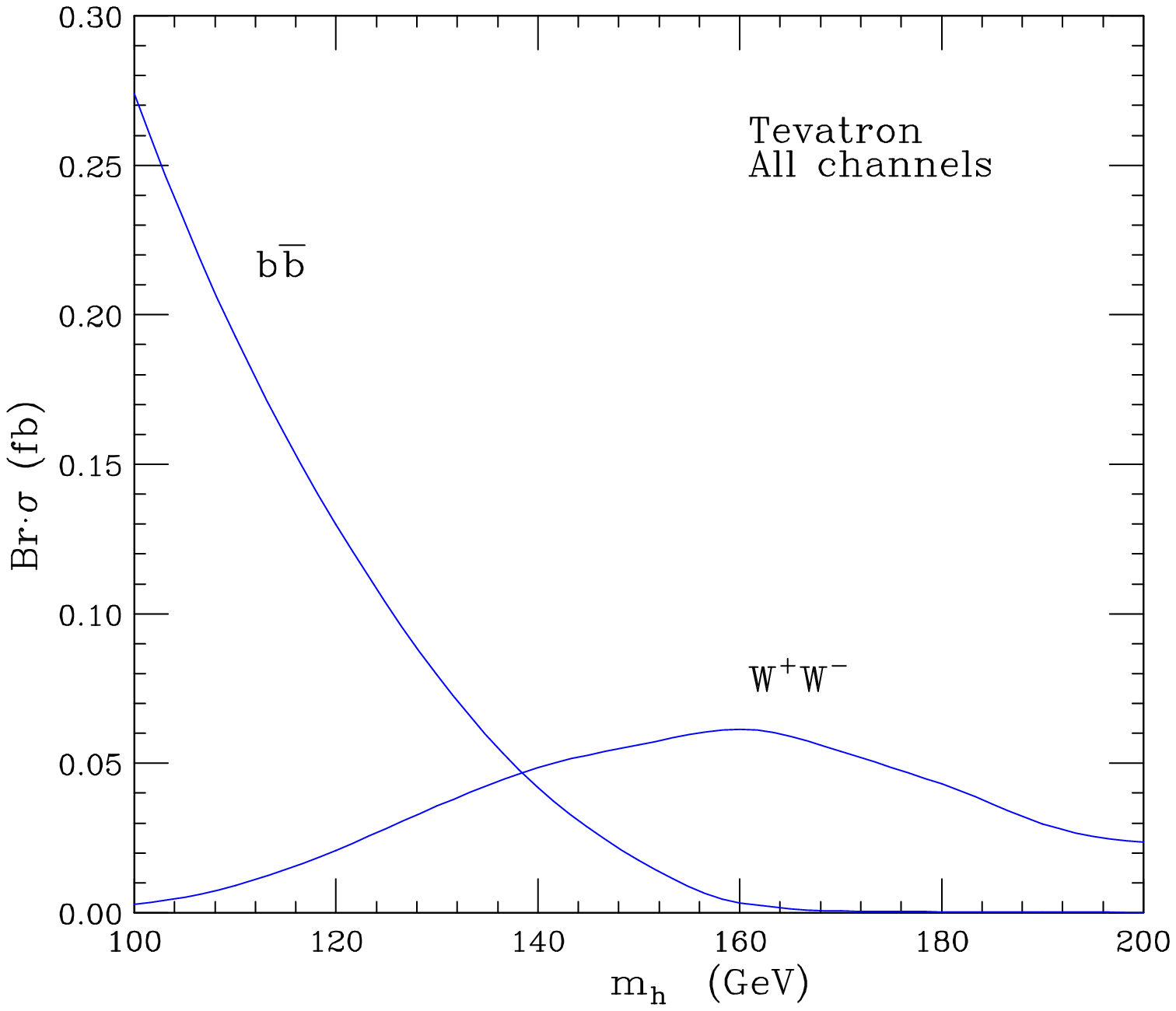}
\vspace*{0cm}
\end{center}
%%%%%%%%%%%%%%%%%%%%%%%%%%%%%%%%%%%%%%%%%%%%%%%%%%%%%%%%%%%%%%%%%%%%%%%%%%%%%
\end{minipage}
\hspace*{-0.5cm}
\begin{minipage}[t]{0.50\textwidth}
%%%%%%%%%%%%%%%%%%%%%%%%%%%%%%%%%%%%%%%%%%%%%%%%%%%%%%%%%%%%%%%%%%%%%%%%%%%%%
\begin{center}
\vspace*{.05cm}
\hspace*{0cm}
\epsfxsize=8cm \epsfbox{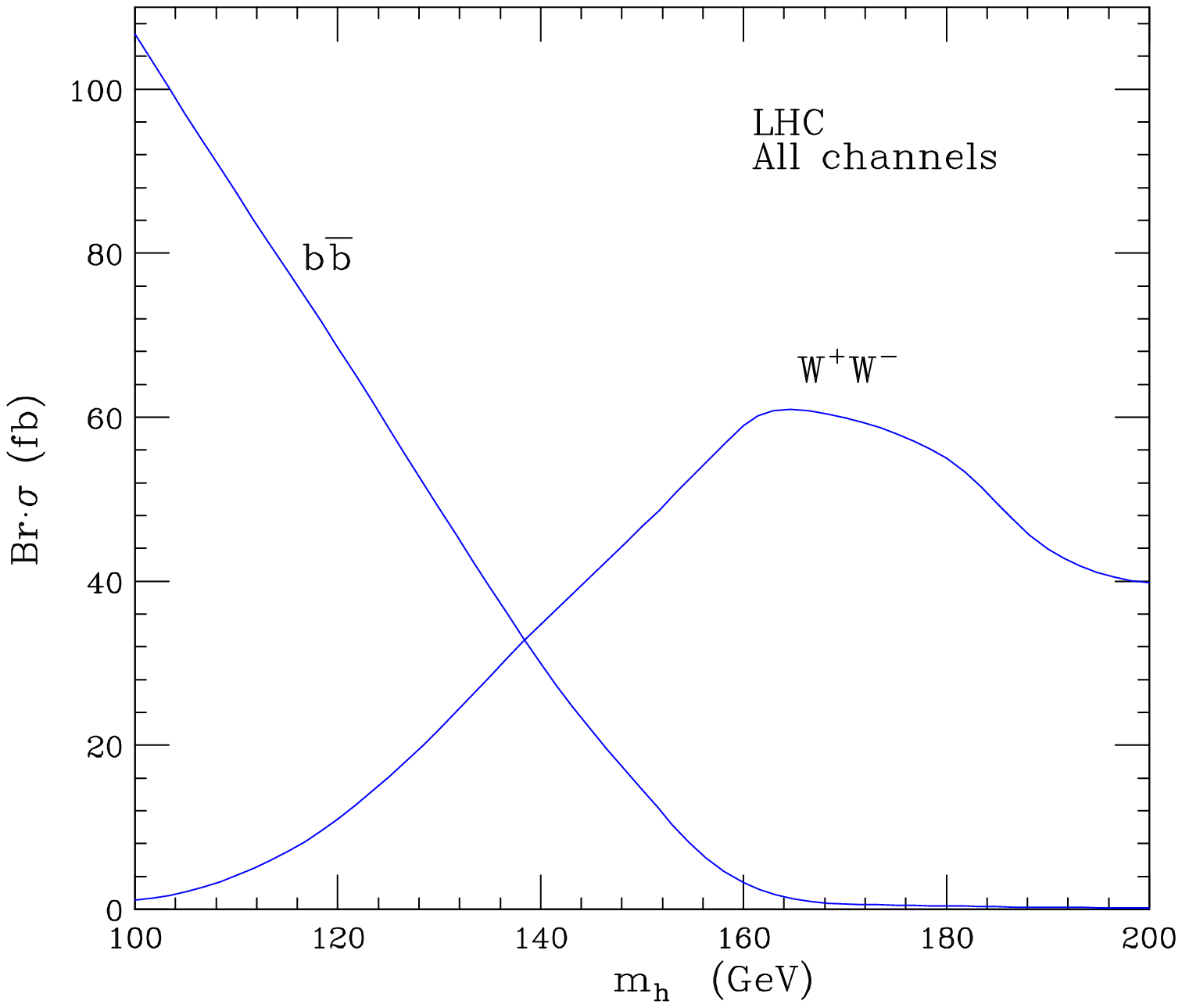}
\vspace*{0cm}
\end{center}
%%%%%%%%%%%%%%%%%%%%%%%%%%%%%%%%%%%%%%%%%%%%%%%%%%%%%%%%%%%%%%%%%%%%%%%%%%%%%
\end{minipage}
\caption{$th$ cross section times the branching ratio of $h \to b\bar{b}$ and
$h \to W^+W^-$ at the Tevatron and at the LHC. The set of parton distribution
functions is CTEQ5L, and the renormalization and factorization scales are set
equal to the Higgs mass.} \label{fig:brxsecs}
\end{figure}

Since the total cross section turns out to be small, detecting any rare decay
of the Higgs, such as $h \to \gamma \gamma$ (whose branching ratio is ${\cal
O}(10^{-3})$),  as suggested in earlier
studies~\cite{Stirling:1992fx,Ballestrero:1993bk,Bordes:1993jy}, is certainly
not feasible. It remains to be seen whether the dominant decay modes of a light
Higgs offer any viable signature. In Fig.~\ref{fig:brxsecs} we show the total
cross section times the branching ratio for $h \to b \bar{b}$ and $h \to W^+
W^- $ (calculated using HDECAY~\cite{Djouadi:1998yw}) at the Tevatron and at
the LHC. The decay into $b\bar{b}$ pairs decreases very quickly and becomes
negligible around Higgs masses of $160$ GeV, exactly where the decay into
$W^+W^-$ reaches its maximum. Since the most challenging mass region for the
Higgs discovery at the LHC is for $m_h \ltap 130$ GeV, we focus our attention
on the Higgs decay into $b\bar{b}$ and fix the Higgs mass to a nominal value of
$115$ GeV.

\begin{figure}[tp]
\begin{center} \vspace*{0cm} \hspace*{0cm} \epsfxsize=5.0cm
\epsfbox{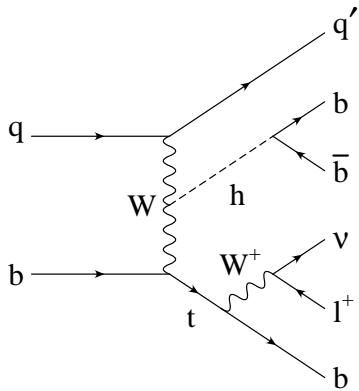}
\vspace*{0cm}
\caption{Example of a Feynman diagram contributing to the signal with three $b$-tags.
The final-state particles are explicitly shown.}
\label{fig:signal_3b}
\end{center}
\end{figure}

We start by presenting the salient kinematic characteristics of the signal,
where the Higgs is required to decay to $b\bar{b}$ and the top to decay
semileptonically ($t \to b \ell^+ \nu$) to provide a hard lepton trigger and
to avoid QCD backgrounds (Fig.~\ref{fig:signal_3b}). We treat the top decay
exactly, including spin and width effects. As in the previous section, we have
chosen the CTEQ5L set of parton distribution functions and fixed the
factorization scale equal to $m_h$.

%%%%%%%%%%%%%%%%%%%%%%%%%%%%%%%%%%%%%%%%%%%%%%%%%%%%%%%%%%%%%%%%%%%%%%%%%%%%%
%%%%%%%%%%%%%%%%%%%%%%%%%%%%%%%%%%%%%%%%%%%%%%%%%%%%%%%%%%%%%%%%%%%%%%%%%%%%%
%%%%%%%%%%%%%%%%%%%%%%%%%%%%%%%%%%%%%%%%%%%%%%%%%%%%%%%%%%%%%%%%%%%%%%%%%%%%
%%%%%%%%%%%%%%%%%%%%%%%%%%%%%%%%%%%%%%%%%%%%%%%%%%%%%%%%%%%%%%%%%%%%%%%%%%%%%
%%%%%%%%%%%%%%%%%%%%%%%%%%%%%%%%%%%%%%%%%%%%%%%%%%%%%%%%%%%%%%%%%%%%%%%%%%%%%
\begin{figure}[tp]
\begin{center}
\vspace*{0cm}
\hspace*{0cm}
\epsfxsize=9cm \epsfbox{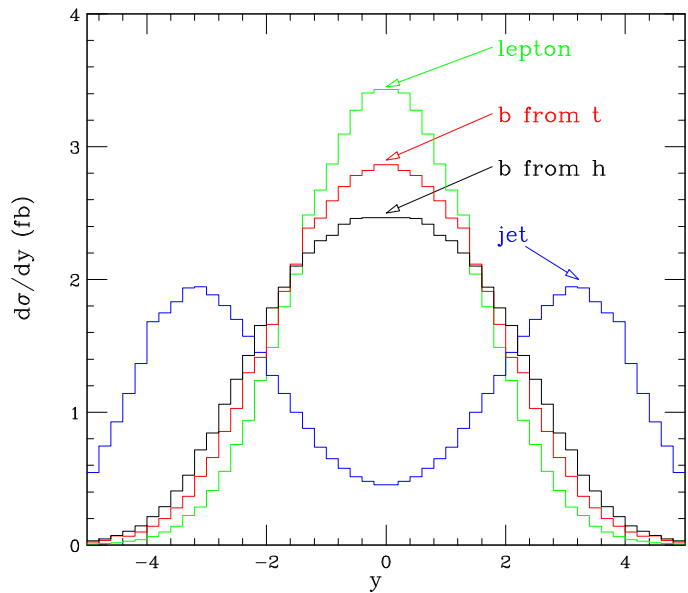}
\vspace*{0cm}
\caption{Rapidity distributions for the final-state particles
(the lepton and the $b$ from the top quark, the $b$'s from the Higgs, and the jet)
in the $t$-channel at the LHC.}
\label{fig:rapidity}
\end{center}
\end{figure}
%%%%%%%%%%%%%%%%%%%%%%%%%%%%%%%%%%%%%%%%%%%%%%%%%%%%%%%%%%%%%%%%%%%%%%%%%%%%%

In Fig.~\ref{fig:rapidity} we show the rapidity distributions of the
final-state particles in the signal events. Both the $b$'s from the Higgs
decay and the $b$ and the lepton from the top decay are produced centrally
while the light quark emitting the virtual $W$ favors large rapidities,
peaking at around 3 units. The presence of a forward jet is related to the
behavior of the cross section as a function of the virtuality of the $W$-boson
exchanged in the $t$-channel, $d\sigma/dq^2 \sim 1/(q^2-M_W^2)^2$. The region
$-q^2\leq M_W^2$ dominates, in analogy to single-top
production~\cite{Willenbrock:1986cr,Yuan:1990tc,Ellis:1992yw}. Since we also
assume that the charge of the $b$-jet is not measured, the signature for this
processes is:
\begin{equation} 3 b + 1\,
{\rm fwd \;jet} + \ell^\pm + \slash \!\!\!p^T\,.
\end{equation}
In order to estimate the number of events in the detector, we have chosen the
acceptances as shown in Table~2, corresponding to low-luminosity running
(${\cal L} = 10^{33}{\rm /cm^2/s}$). With 30 fb$^{-1}$ we expect around $120$ events.
When the $b$-tagging efficiency ($\epsilon_{b }=60 \% $) and lepton efficiency
($\epsilon_{\ell }=90 \% $) are included the number of expected events goes
down to $23$.\footnote{The efficiencies are taken from Ref.~\cite{TDRs}.}
Although the final tally is low, this is more than half of the number of
events expected for the $t\bar{t} h$ process after branching ratios and
reconstruction efficiencies are taken into account~\cite{Beneke:2000hk}.
However, the impact of the backgrounds is more severe for Higgs plus
single top, as we discuss in the following.

% Irreducible backgrounds

The largest sources of irreducible background are from single-top production
in association with a $b\bar{b}$ coming either from the resonant production of
a $Z$ boson ($tZ$) or from a higher-order QCD process, such as the emission of
a gluon subsequently splitting into a $b\bar{b}$ pair ($tb\bar{b}$). Although
the final-state particles in the above processes are exactly the same as in
the signal, the typical invariant mass $m_{b\bar{b}}$ of the $b$'s in the
final state is quite different.
%%%%%%%%%%%%%%%%%%%%%%%%%%%%%%%%%%%%%%%%%%%%%%%%%%%%%%%%%%%%%%%%%%%%%%%%%%%%%%%
\begin{table}[p]
\label{tab:detectorcuts}
\caption{Cuts applied to the $t$-channel signal at the
LHC (low luminosity), with three and four $b$-tags, for $m_h=115\; {\rm GeV}$.
The values of the cross sections after the cuts are applied are shown
in the last two columns. Branching ratios Br$(h\to b\bar{b})=77\%$
as well as Br$(W \to \ell \nu) =22\%$ are included.
Detector efficiencies are not included.}
\medskip
\addtolength{\arraycolsep}{0.1cm}
\renewcommand{\arraystretch}{1.4}
\begin{center}
\begin{tabular}[4]{ccccccc|cc}
\hline\hline cut                & $p^{T}_{b}>$       & $p^T_{\ell,\nu}>  $&
$p^T_{j}>         $& $|\eta_{b,\ell}|< $& $|\eta_j|       < $& $\Delta
R_{ij}   >$& $\sigma_{3b}      $&
$\sigma_{4b} $      \\
\hline value & $15$ GeV & $20$ GeV& $30$ GeV & $2.5$ &  $5$ & $0.4$ & 4.0  fb &
1.9 fb \\
 \hline
 \hline
\end{tabular}
\end{center}
\end{table}
%%%%%%%%%%%%%%%%%%%%%%%%%%%%%%%%%%%%%%%%%%%%%%%%%%%%%%%%%%%%%%%%%%%%%%%%%%%%%%%%%%%%%
\nopagebreak[4]
\begin{table}[p]
\label{tab:backgrounds3b}

\caption{Cross sections (fb) for the signal and some of the most important
backgrounds for Higgs plus single-top production in the $t$-channel at the LHC
(low luminosity), with three $b$-tags, for $m_h=115\; {\rm GeV}$. Branching
ratios into final states are included, as well as the $b$-tagging efficiency
$\epsilon_b=60\%$ and the lepton-tagging efficiency $\epsilon_\ell=90\%$. The
backgrounds include both the irreducible ones ($tZ$ and $tb\bar{b}$) and the
reducible ones ($t\overline{t}$ and $t\overline{t}j$). In the reducible
backgrounds, a $c$ quark from the decay of a $W$ is mistagged as a $b$ quark
(the mistag probability, $\epsilon_{c }=10 \% $, is included). ``Detector
cuts'' correspond to the choice of cuts in Table~2. In the second line,
assuming the top is correctly reconstructed, the invariant mass of the other
two $b$'s is required to be in a window of $m_h \pm 22$ GeV ($95\%$ of the
signal and $40\%$ of the $tZ$ background is assumed to fall in this range). In
the third line, a forward jet tag is added. In the fourth line a minimum
invariant mass of $250$ GeV for the Higgs candidate and the forward jet is
required. The last line gives the expected number of events with 30 fb$^{-1}$
of integrated luminosity at the LHC. }

\addtolength{\arraycolsep}{0.1cm}
\renewcommand{\arraystretch}{1.4}
\medskip
\begin{center} \begin{tabular}[4]{c|ccccc}
\hline
\hline
& & \multicolumn{3}{c}{3$b$-tag (low luminosity)}& \\[1pt]
& Signal &  $tZ$  & $tb\bar{b}$  & $t\overline{t}$ & $t\overline{t}j$ \\
\hline
Detector cuts                 &  0.80   & 2.1  &  4.1  & 810 &   100 \\[7pt]
$|m_{b\bar{b}}-m_h|<22$ GeV   &  0.75   & 0.83 &  0.54 & 450 &    38 \\[7pt]
$|\eta_j|>2, p^T_j>50$ GeV    &  0.39   & 0.44 &  0.26 &  13 &   8.0 \\[7pt]
$m_{b\bar bj}>250$ GeV        &  0.35   & 0.35 &  0.25 &   - &   7.4 \\[7pt]
\hline
Events with 30 fb$^{-1}$      &  10     & 10   &  7    &   - &   220 \\[3pt]
 \hline
\hline
\end{tabular}

\end{center}
\end{table}
%%%%%%%%%%%%%%%%%%%%%%%%%%%%%%%%%%%%%%%%%%%%%%%%%%%%%%%%%%%%%%%%%%%%%%%%%%%%%%%%%%%%%
Let us study the idealized case where the $t$ is reconstructed with 100\%
efficiency, such that we know which $b$ comes from top decay. For $tZ$ the
distribution in $m_{b\bar{b}}$ is peaked around the $Z$ mass, while for
$tb\bar{b}$ it is largest at small invariant mass. We require that the
invariant mass of the $b\bar{b}$ pair lies in a window $m_h \pm 2 \sigma$,
where $\sigma=11$ GeV is the expected experimental resolution~\cite{TDRs}.
Assuming a Gaussian distribution, we estimate that $40\%$ of the events coming
from $tZ$ fall in this range (for $m_h=115$ GeV), decreasing quickly for larger
Higgs masses.  The cross sections for the signal and these two irreducible
backgrounds are given in Table~3 with the cut on the invariant mass of the
$b\bar b$ applied (second row).  We see that the backgrounds are comparable to
the signal after this cut.

% Reducible backgrounds

\begin{figure}[tbp]
\begin{minipage}{1\textwidth}
\begin{center}

\vspace*{0cm} \hspace*{0cm} \epsfxsize=10cm
\epsfbox{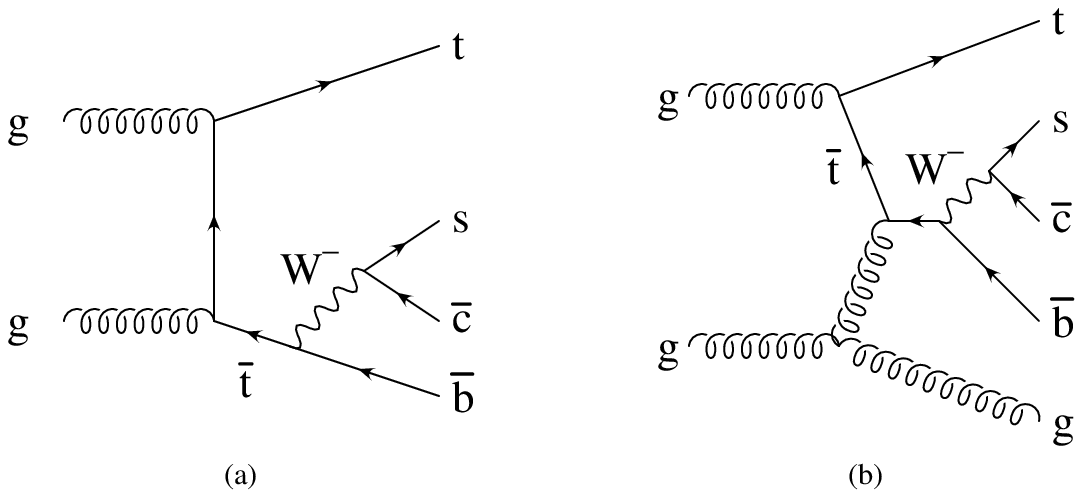} \vspace*{0cm} \caption{Reducible backgrounds
in the $3b$-tag analysis coming from the production of a $t\bar{t}$ pair and
jets. The $c$ quark coming from the decay of a $W$ is misidentified as a $b$
quark. In $t\bar{t}$ production (a) the $s$ quark is the forward jet while in
$t\bar{t}j$ production (b) the  $s$-quark jet is missed.   }
\label{fig:ttbar_backgrounds}
%%%%%%%%%%%%%%%%%%%%%%%%%%%%%%%%%%%%%%%%%%%%%%%%%%%%%%%%%%%%%%%%%%%%%%%%%%%%%
\vspace*{0.2cm} \hspace*{0cm} \epsfxsize=5cm
\epsfbox{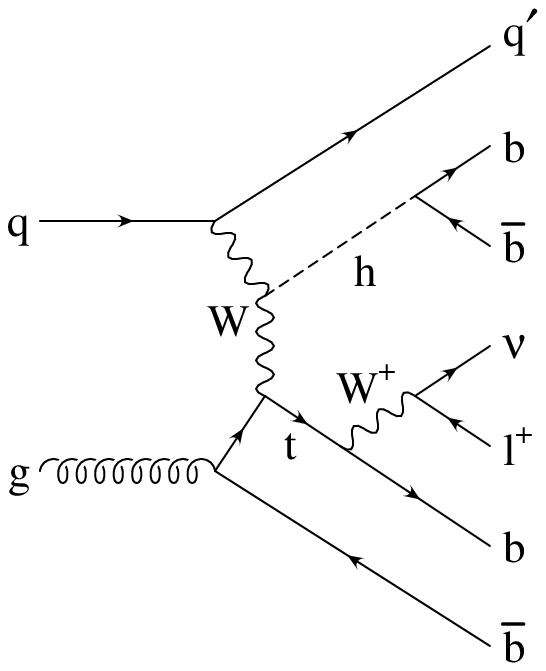}
\vspace*{0cm}
\caption{Example of a Feynman diagram
contributing to the signal in the $4b$-tag analysis. }
\label{fig:t-channel_gluon}
%%%%%%%%%%%%%%%%%%%%%%%%%%%%%%%%%%%%%%%%%%%%%%%%%%%%%%%%%%%%%%%%%%%%%%%%%%%%%
\vspace*{0.7cm} \hspace*{0cm} \epsfxsize=4.8cm
\epsfbox{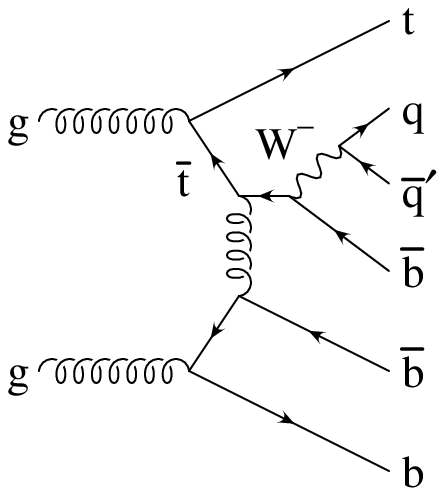}
\vspace*{-.2cm}
\caption{Reducible background in the $4b$-tag analysis coming from
the production of $t\bar{t}b\bar{b}$. One of the quarks coming from the
$W$ is missed while the other provides the forward tag. }
\label{fig:ttbb}

\end{center}
\end{minipage}
\end{figure}

An important reducible background comes from the production of a $t\bar{t}$
pair [with $t\bar{t}\to (W^+ \to \ell^+ \nu)  (W^- \to \bar{c} s) b\bar{b} $],
as shown in Fig.~\ref{fig:ttbar_backgrounds}(a) (fourth column of
Table~3).\footnote{ Other sources of reducible background come from the
production of a $W$ in association with four jets of which three are (or are
misidentified as) $b$ quarks.} This process contributes to the background when
the $c$ quark coming from the hadronic decay of one of the $W$'s is
misidentified as a $b$ quark and the $s$ quark is the forward jet. A mistag
probability $\epsilon_{c }=10\%$ is included in the cross sections quoted in
Table~3.\footnote{The mistag probability quoted in Ref.~\cite{TDRs} is
$\epsilon_{c }=14\%$, but no specific effort was made to minimize it. We assume
that it can be reduced to $10\%$ while maintaining high $b$-tagging
efficiency.} Even in the idealized case where one top quark is reconstructed
with 100\% efficiency, the number of background events is very large.  This
background is drastically reduced by requiring the presence of the forward jet
(third row of Table~), but it is still large compared with the signal. To
reduce this background further one can exploit the fact that the forward jet
and the $bc$ that fake the Higgs signal all come from top decay, so their
invariant mass is nominally 175 GeV.  We therefore require that the invariant
mass of the forward jet and the $b\bar b$ pair exceed 250 GeV (fourth column of
Table~3).  This essentially eliminates the $t\bar t$ background,\footnote{In
actuality some of the background will pass the cut due to jet resolution.}
while maintaining most of the signal.

There is a related background, $t\bar tj$ [shown in
Fig.~\ref{fig:ttbar_backgrounds}(b)], of which one cannot so easily dispose
(fifth column of Table~3). In this case the amplitude is dominated by the
exchange of a gluon in the $t$-channel and the jet is naturally produced
forward, while both top quarks remain central. If the $s$-quark jet is missed,
the distributions of the remaining particles (the $b$'s, the mistagged $c$
quark, and the lepton) are very similar to the ones in Fig.~\ref{fig:rapidity}.
After all cuts are applied, the number of background events is large compared
with the signal. We conclude that at the LHC the measurement of Higgs plus
single top with three $b$-tags is hampered by the overwhelming $t\bar{t}j$
background.

%%%%%%%%%%%%%%%%%%%%%%%%%%%%%%%%%%%%%%%%%%%%%%%%%%%%%%%%%%%%%%%%%%%%%%%%%%%%%
%%%%%%%%%%%%%%%%%%%%%%%%%%%%%%%%%%%%%%%%%%%%%%%%%%%%%%%%%%%%%%%%%%%%%%%%%%%%%

%4-b tag case

Another possibility for reducing the background is to consider four $b$-tags (see
Fig.~\ref{fig:t-channel_gluon}). Since the $b$ distribution in the proton sea
arises from the splitting of virtual gluons into collinear $b\bar{b}$ pairs,
the additional $b$ tends to reside at small $p^T$. However, some fraction of
the time this additional $b$ will be at high $p^T$ and be detected. Studies
performed on single-top production have shown that $p^T_{\rm min}>15$ GeV
needed at the LHC to detect a jet is enough for the perturbative calculation
to be reliable~\cite{Stelzer:1998ni}. The $4b$-tag case can be analyzed along
the same lines as above. When detector acceptance is taken into account, the
cross section is around one half of the $3b$-tag one (last column in Table~2).
Both irreducible and reducible backgrounds are present. The irreducible
backgrounds are analogous to $tZ$ and $tb\bar{b}$ discussed in the $3b$-tag
case, where an additional $b$ present in the final state (arising, as in the
signal, from an initial gluon splitting into $b\bar b$) is also detected. We
again assume that the top quark is reconstructed with 100\% efficiency,
leaving three pairs of $b$'s in the final state that could have come from
Higgs decay. We give in Table~4 the cross sections with
detector cuts and with the requirement that the invariant mass of at least one
$b\bar b$ pair lies in a window $m_h \pm 22$ GeV.  A forward jet cut is added
in the third row of Table~4, and a requirement that the
minimum invariant mass of all $b\bar b$ pairs (excluding the $b$ from top
decay) exceed 90 GeV in the fourth row.  This last cut reduces the $tb\bar
b(b)$ background, because the $b\bar b$ pair, which comes from
gluon-splitting, tends to reside at low invariant mass.  After all cuts, the
irreducible backgrounds are comparable to the signal.

There are several reducible backgrounds to consider, all with top pairs in the
final state.  We give in the fourth column of Table~4
the cross section for $t\bar{t} b \bar{b}$.  This process contributes through
the decay $t\bar{t} b \bar{b} \to W^+W^- b \bar{b} b \bar{b} $, where one $W$
decays hadronically to two jets, one of which is identified as the forward jet
while the second is missed (Fig.~\ref{fig:ttbb}).  The forward jet cut and the
minimum $b\bar b$ mass cut reduce this background to the same level as the
signal.  A related background, given in the fifth column of
Table~4, occurs when the hadronically-decaying $W$ yields
a (mistagged) charm quark.  Of the remaining quarks (one $s$ and three $b$'s),
either the $s$ or one of the $b$'s provides the forward jet, and one is missed.
The cuts similarly reduce this background to the same level as the signal.
There is also a background from $t\bar tj$, where the hadronically-decaying $W$
yields a $c$ and $s$ quark, both of which are mistagged ($\epsilon_s =
1\%$). This background is the largest of all, but it is removed by the
requirement on the minimum $b\bar b$ invariant, since the (mistagged) $cs$
pair comes from $W$ decay.\footnote{In actuality, some of this background will
remain due to jet resolution.}

%%%%%%%%%%%%%%%%%%%%%%%%%%%%%%%%%%%%%%%%%%%%%%%%%%%%%%%%%%%%%%%%%%%%%%%%%%%%%%%%%%%%%

\nopagebreak[4]
\begin{table}[p]
\label{tab:bkg4b}

\caption{Cross sections (fb) for the signal and some of the most
important backgrounds for Higgs plus single-top production in the
$t$-channel at the LHC (low luminosity), with four $b$-tags, for
$m_h=115\; {\rm GeV}$. Branching ratios into final states are
included, as well as the $b$-tagging efficiency $\epsilon_b=60\%$
and the lepton-tagging efficiency $\epsilon_\ell=90\%$. The
backgrounds include both the irreducible ones [$tZ(b)$ and
$tb\bar{b}(b)$] and the reducible ones [$t\overline{t}b
\overline{b}$, $t\overline{t}b \overline{b}$ (mistag), $t\bar
tj$]. In $t\overline{t}b \overline{b}$ (mistag) and $t\bar tj$, a
$c$ quark from the decay of a $W$ is mistagged as a $b$ quark (the
mistag probability, $\epsilon_{c}=10 \% $, is included); in $t\bar
tj$, an $s$ quark from the decay of a $W$ is mistagged (the mistag
probability, $\epsilon_s=1\%$, is included). ``Detector cuts''
correspond to the choice of cuts in Table~2. In the second line,
assuming the top is correctly reconstructed, the invariant mass of
at least one pair of the other three $b$'s is required to be in a
window of $m_h \pm 22$ GeV ($95\%$ of the signal and $40\%$ of the
$tZ$ background is assumed to fall in this range). In the third
line, a forward jet tag is added. In the fourth line a minimum
invariant mass of $90$ GeV for all $b \bar b$ pairs (not including
the $b$ that reconstructs the top quark) is required. The last
line gives the expected number of events with 30 fb$^{-1}$ of
integrated luminosity at the LHC.}

\addtolength{\arraycolsep}{0.1cm}
\renewcommand{\arraystretch}{1.4}
\medskip
\begin{center} \begin{tabular}[4]{c|cccccc}
\hline \hline
& & \multicolumn{4}{c}{4$b$-tag (low luminosity)}\\[1pt]

& Signal &  $tZ(b)$  & $tb\bar{b}(b)$  & $ t\overline{t}b
\overline{b}$ &
$t\bar tb\bar b$ (mistag) & $t\overline{t}j$ \\
\hline
Detector cuts                 & 0.22& 0.42 & 1.5  & 5.8   & 3.1  & 9.0   \\[7pt]
$|m_{b\bar{b}}-m_h|<22$ GeV   & 0.21& 0.17 & 0.61 & 2.6   & 2.3  & 6.3   \\[7pt]
$|\eta_j|>2$                  & 0.15& 0.11 & 0.41 & 0.17  & 0.18 & 2.4   \\[7pt]
min $m_{b\bar b}>90$ GeV      & 0.1 & 0.065& 0.08 & 0.053 & 0.078&  -    \\[7pt]
\hline
Events with 30 fb$^{-1}$      & 3.0 &  1.9  & 2.5  & 1.6  & 2.3  &  -    \\[3pt]
 \hline
\hline
\end{tabular}

\end{center}

\end{table}

%%%%%%%%%%%%%%%%%%%%%%%%%%%%%%%%%%%%%%%%%%%%%%%%%%%%%%%%%%%%%%%%%%%%%%%%%%%%%%%%%%%%%
%%%%%%%%%%%%%%%%%%%%%%%%%%%%%%%%%%%%%%%%%%%%%%%%%%%%%%%%%%%%%%%%%%%%%%%%%%%%%%%%%%%%%

\nopagebreak[4]
\begin{table}[p]
\label{tab:bkg4bhighlum}

\caption{Cross sections (fb) for the signal and some of the most
important backgrounds for Higgs plus single-top production in the
$t$-channel at the LHC (high luminosity), with four $b$-tags, for
$m_h=115\; {\rm GeV}$.  Branching ratios into final states are
included, as well as the $b$-tagging efficiency $\epsilon_b=50\%$
and the lepton-tagging efficiency $\epsilon_\ell=90\%$. The
backgrounds include both the irreducible ones [$tZ(b)$ and
$tb\bar{b}(b)$] and the reducible ones [$t\overline{t}b
\overline{b}$, $t\overline{t}b \overline{b}$ (mistag), $t\bar
tj$]. In $t\overline{t}b \overline{b}$ (mistag) and $t\bar tj$, a
$c$ quark from the decay of a $W$ is mistagged as a $b$ quark (the
mistag probability, $\epsilon_{c}=10 \% $, is included); in $t\bar
tj$, an $s$ quark from the decay of a $W$ is mistagged (the mistag
probability, $\epsilon_s=1\%$, is included). ``Detector cuts''
correspond to the choice of cuts in Table~2, apart from the
minimum $p^T_b$ which is now raised to $30$ GeV. In the second
line, assuming the top is correctly reconstructed, the invariant
mass of at least one pair of the other three $b$'s is required to
be in a window of $m_h \pm 22$ GeV ($95\%$ of the signal and
$40\%$ of the $tZ$ background is assumed to fall in this range).
In the third line, a forward jet tag is added. In the fourth line
a minimum invariant mass of $90$ GeV for all $b \bar b$ pairs (not
including the $b$ that reconstructs the top-quark) is required.
The last line gives the expected number of events with 300
fb$^{-1}$ of integrated luminosity at the LHC.}

\addtolength{\arraycolsep}{0.1cm}
\renewcommand{\arraystretch}{1.4}
\medskip
\begin{center} \begin{tabular}[4]{c|cccccc}
\hline \hline
& & \multicolumn{4}{c}{4$b$-tag (high luminosity)}\\[1pt]

& Signal &  $tZ(b)$  & $tb\bar{b}(b)$  & $ t\overline{t}b
\overline{b}$ &
$t\bar tb\bar b$ (mistag) & $t\overline{t}j$ \\
\hline
Detector cuts                 & 0.061& 0.094 & 0.23  & 4.0   & 1.5  & 3.3   \\[7pt]
$|m_{b\bar{b}}-m_h|<22$ GeV   & 0.058& 0.037 & 0.096 & 1.7   & 1.1  & 2.5   \\[7pt]
$|\eta_j|>2$                  & 0.040& 0.025 & 0.067 & 0.15  & 0.11 & 0.94  \\[7pt]
min $m_{b\bar b}>90$ GeV      & 0.032& 0.018 & 0.027 & 0.069 & 0.068&  -    \\[7pt]
\hline
Events with 300 fb$^{-1}$     & 9.5 &  5.5   & 8.0   & 21    & 21   &  -    \\[3pt]
 \hline
\hline
\end{tabular}

\end{center}
\end{table}
%%%%%%%%%%%%%%%%%%%%%%%%%%%%%%%%%%%%%%%%%%%%%%%%%%%%%%%%%%%%%%%%%%%%%%%%%%%%%

Although each background in the $4b$-tag analysis is comparable to the signal,
there are only a few signal events with 30 fb$^{-1}$. Therefore, there is little
hope of observing a signal in this channel, unless significantly
more than 30 fb$^{-1}$ can be delivered while maintaining the same detector
performance.  At high luminosity (${\cal L}=10^{34}/{\rm cm^2/s}$), it is
anticipated that the minimum $p_T$ for jets must be raised to 30 GeV. In
Table~5 we study the signal and backgrounds in
this scenario (the $b$-tagging efficiency is also lowered to 50\%).  After all
cuts, the $t\bar tb\bar b$ backgrounds are now each twice as large as the
signal, because these backgrounds involve missing a jet, which is more likely
with the increased jet $p_T$ threshold.  The number of signal events in 300
fb$^{-1}$ is about 10, with about 55 background events.  Significantly more
integrated luminosity would be needed to see a signal in this channel.

%%%%%%%%%%%%%%%%%%%%%%%%%%%%%%%%%%%%%%%%%%%%%%%%%%%%%%%%%%%%%%%%%%%%%%%%%%%%%
\begin{figure}[tbp]
\begin{minipage}[t]{0.50\textwidth}
%%%%%%%%%%%%%%%%%%%%%%%%%%%%%%%%%%%%%%%%%%%%%%%%%%%%%%%%%%%%%%%%%%%%%%%%%%%%%
\begin{center}
\vspace*{0cm}
\hspace*{0cm}
\epsfxsize=8.1cm \epsfbox{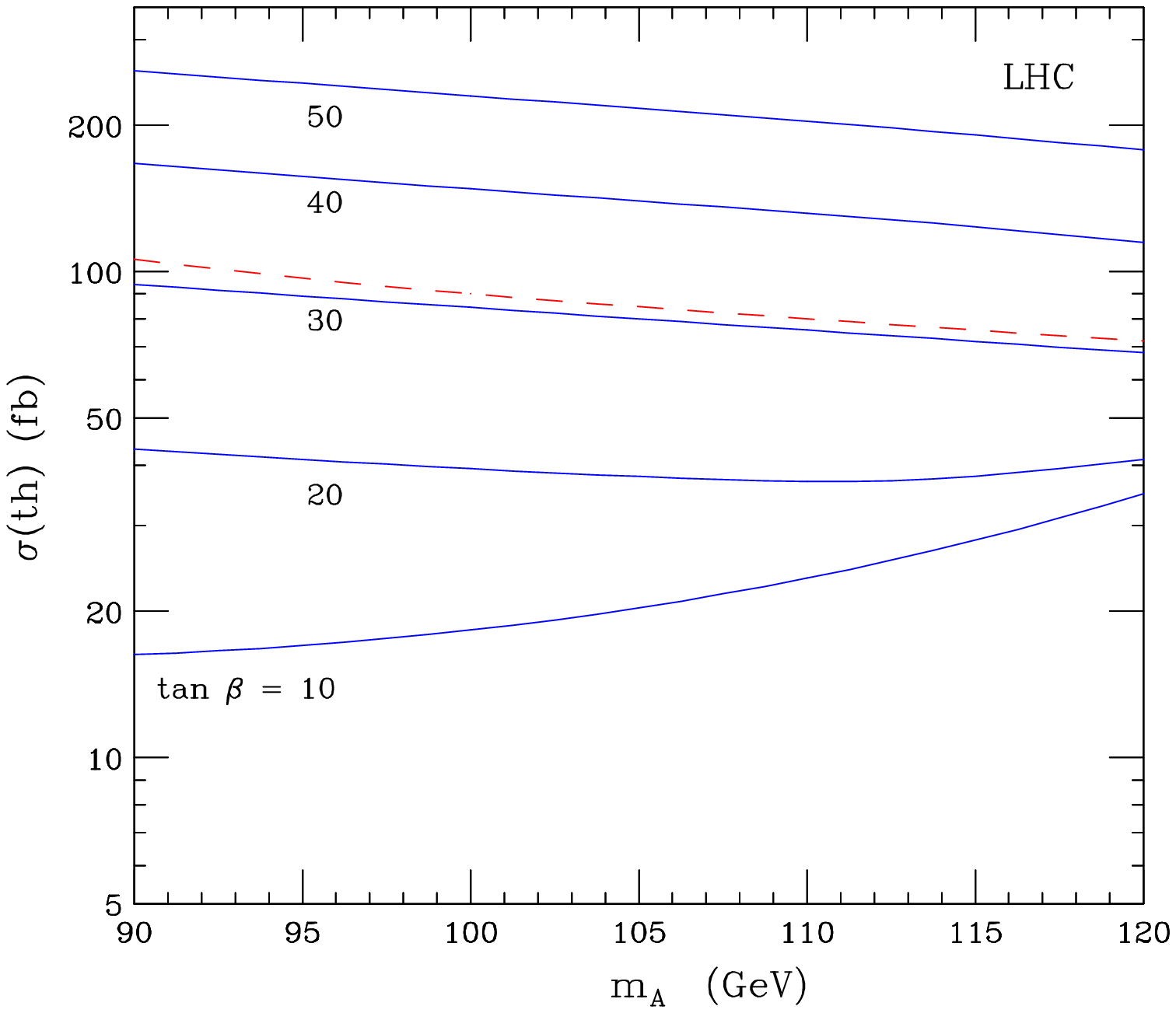}
\vspace*{0cm}
\end{center}
%%%%%%%%%%%%%%%%%%%%%%%%%%%%%%%%%%%%%%%%%%%%%%%%%%%%%%%%%%%%%%%%%%%%%%%%%%%%%
\end{minipage}
\hspace*{-0.3cm}
\begin{minipage}[t]{0.50\textwidth}
%%%%%%%%%%%%%%%%%%%%%%%%%%%%%%%%%%%%%%%%%%%%%%%%%%%%%%%%%%%%%%%%%%%%%%%%%%%%%
\begin{center}
\vspace*{.05cm}
\hspace*{0cm}
\epsfxsize=8.1cm \epsfbox{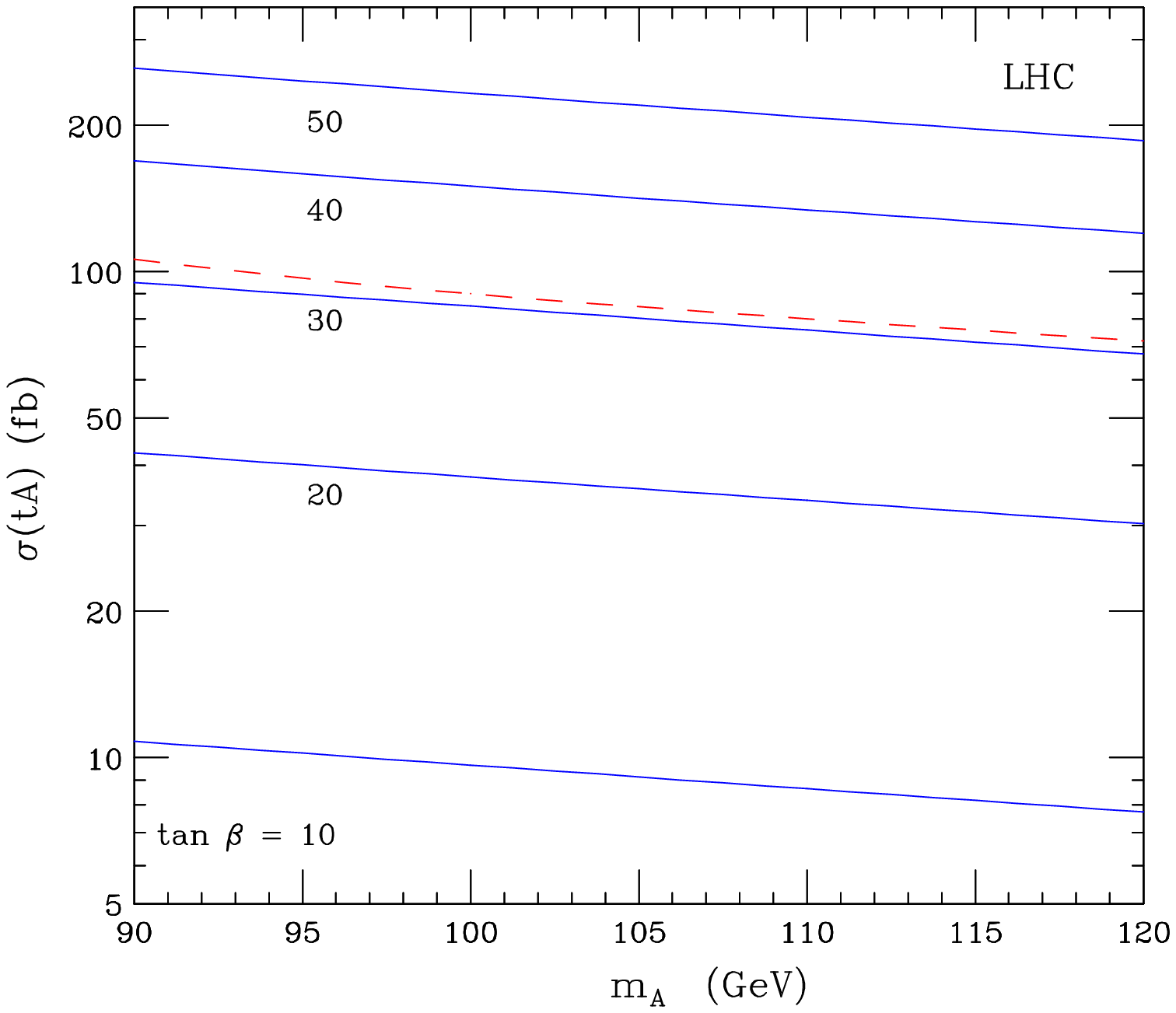}
\vspace*{0cm}
\end{center}
%%%%%%%%%%%%%%%%%%%%%%%%%%%%%%%%%%%%%%%%%%%%%%%%%%%%%%%%%%%%%%%%%%%%%%%%%%%%%
\end{minipage}
\caption{Cross sections for production of CP-even Higgs $h$ and CP-odd
 Higgs $A$ in association with single top as a function
 of $m_A$ and $\tan \beta$ ($M_{\rm SUSY}= 1$ TeV, $\mu=-200$ GeV and maximal
 stop-squark mixing is assumed). Only $t$-channel production is included.
 The cross section for a standard-model Higgs with
 $m_{h_{\rm SM}}=m_A$ is given as a reference (dashes). The set of parton distribution functions is CTEQ5L
 and the factorization scale is set equal to the Higgs mass.}
\label{fig:xsecsSUSY}
\end{figure}
%%%%%%%%%%%%%%%%%%%%%%%%%%%%%%%%%%%%%%%%%%%%%%%%%%%%%%%%%%%%%%%%%%%%%%%%%%%%%

%%%%%%%%%%%%%%%%%%%%%%%%%%%%%%%%%%%%%%%%%%%%%%%%%%%%%%%%%%%%%%%%%%%%%%%%%%%%%%%%%%%%%

\section{Production of supersymmetric Higgs bosons}
\label{sec:SUSY}

It is interesting to ask whether there could be an enhancement in the signal
when the production of non-minimal Higgs  bosons is considered. With this aim
we have investigated the production of a light CP-even ($h$) and a CP-odd
($A$) Higgs in the MSSM.

The Higgs sector of the MSSM is the same as the 2HDM presented in Appendix A
except that it depends (at tree-level) on only
two free parameters, which can be chosen to be $m_A$ and $\tan \beta$.
The tree-level relations between the Higgs masses are modified by
radiative corrections that involve the supersymmetric particle spectrum,
mainly of the top sector~\cite{Haber:1991aw,Okada:1991vk,Ellis:1991nz}. Since
the analytical form of the corrections is quite involved (see
Ref.~\cite{Haber:1997fp}) we used HDECAY~\cite{Djouadi:1998yw} to evaluate the
Higgs-boson masses and the mixing parameter $\alpha$, given $m_A$, $\tan
\beta$ and information on the stop-quark mixings and masses.

For large $m_A$, the masses of the heavy Higgs particles approximately
coincide, $m_A\simeq m_H \simeq m_{H^\pm}$, while the CP-even Higgs remains
light. This is the so-called decoupling limit where the standard model
couplings and particle content are recovered. In the case of large $\tan \beta$
and small $m_A$ one finds that $ m_h\simeq m_A$ and the Higgs couplings to the
vector bosons and to the fermions are different from those predicted by the
standard model. In particular, there is a strong enhancement of the
bottom-quark coupling to both the $h$ and the $A$, which can give rise to
interesting signatures at the
colliders~\cite{Carena:2000yx,Dicus:1989cx,Dai:1995vu,Dai:1996rn}. We focus our
attention in this area of the parameter space, which is not excluded by the
measurements from LEP~\cite{LEPEWWG}, choosing $m_A<120$ GeV and $10< \tan
\beta<50$.

In Fig.~\ref{fig:xsecsSUSY} we show the cross section for production of the
CP-even Higgs $h$ and CP-odd Higgs $A$ in association with single top as a
function of $m_A$ and $\tan \beta$. These are calculated using tree-level
matrix elements generated by MADGRAPH~\cite{Stelzer:1994ta} (and checked
against those obtained by COMPHEP~\cite{Pukhov:1999gg}) convoluted with the
parton distribution function set CTEQ5L~\cite{Lai:1995bb}, and with the
renormalization and factorization scales set equal to the Higgs mass. We
assume a simplified scenario where the third generation diagonal
soft-supersymmetry-breaking squark masses are degenerate, with common value
$M_{\rm SUSY}=1$ TeV, and the mixing between the top-squarks is maximal,
$X_t=A_t-\mu \cot \beta=\sqrt{6}M_{\rm SUSY}$, with $\mu=-200$ GeV (for an
extensive discussion on the other possible choices see
Ref.~\cite{Carena:2000yx} and references therein).

As shown in Fig.~\ref{fig:xsecsSUSY}, for $\tan \beta \gtap 30$ the cross sections are
indeed enhanced with respect to that for a standard-model Higgs. However, the
increase is never very large. This is basically due to two reasons. First,
from the arguments presented in Section 2 and Appendix A, unitarity imposes
large cancellations among the various diagrams, even in the MSSM Higgs sector.
In this respect, the production of the CP-odd state $A$ is particularly
instructive. Due to its CP quantum numbers, this state cannot couple to two
$W$'s and therefore the contribution from the second diagram in
Fig.~\ref{fig:t-channel} vanishes. One might guess that the destructive
interference with the diagrams where $A$ couples to the quarks cannot take
place anymore and the signal could be much larger. In fact, the complete
calculation shows that the diagram where the $A$ couples to the $W$ and a
charged Higgs $H^+$ (see the second diagram in Fig.~\ref{fig:t-chan-2HDM}) provides the terms
which cancel the large (and unitarity-violating) contributions coming from the
other diagrams (Appendix A). Second, the effects due to the choice of a large
value of $\tan \beta$ work in opposite directions for the bottom and the top
quark, leading to an enhancement of the coupling of the Higgs to bottom but to
a suppression for the top quark. As a result the rates for the $h$ and
the $A$ are comparable to that of a standard-model Higgs with a similar mass for
$m_b \tan \beta \approx m_t$.
For instance, taking $m_h=m_A=115$ GeV and $\tan \beta =50$, we have
$\sigma(th) \simeq \sigma(tA)=190$ fb, which is 2.5 times the cross section
expected in the standard model. Considering the production of the two Higgs
bosons together,\footnote{There is no interference between the two processes due
to the different CP properties of the Higgs bosons.} it would be possible to
achieve a significance ${S}/{\sqrt{B}}\simeq 5$ in the $4b$-tag analyses (see
Tables~4 and 5).

\section{Conclusions}
\label{sec:Conclusions}

In this paper we revisited the production of the Higgs boson in association with single
top at hadron colliders. We provided the full set of cross sections at both the
Tevatron and the LHC for the three production processes ($t$-channel,
$s$-channel and $W$-associated) and we investigated in some detail why they
are smaller than what one would expect comparing with $t\bar t h$ production.
For the $t$-channel, which gives the most important contribution at the LHC,
this is due to large cancellations taking place between different diagrams. We
have shown that the above peculiarity is not accidental but is a consequence
of the renormalizability of theory, and we gave a detailed proof in the
general framework of a two-Higgs-doublet model.

Focusing on the $t$-channel process, we discussed the possibility
of detecting the production of Higgs plus single top at the LHC,
concentrating on the decay of the Higgs into $b \bar b$.  We
considered events where three and four $b$ quarks are tagged.  In
the case of three $b$-tags, there is an overwhelming background from
$t\bar tj$.  In the case of  four $b$-tags there is no single
overwhelming background, but rather several backgrounds that are
comparable to the signal. Given our present expectations for
detector capabilities and luminosity at the LHC, it seems unlikely
that one can extract a signal from the backgrounds.

There are several things that could improve this prognosis.
Several of the backgrounds involve a mistagged $c$  quark,
and if the mistag rate  can be reduced
significantly below 10\%, these backgrounds
would be less severe.  One might also be able to find a more
efficient set of cuts to reduce the backgrounds. Since the signal
involves  the $t$-channel exchange of a $W$ boson, one might be
able to use a rapidity gap to distinguish the signal from the
reducible backgrounds (the irreducible backgrounds also involve
$t$-channel $W$ exchange, however)~\cite{Rainwater:1996ud}.

Finally we have also presented the results for the $t$-channel production of
the CP-even state $h$ and the CP-odd state $A$ of the MSSM at the LHC. For
$m_A<120$ GeV and large $\tan \beta$ there is a moderate enhancement of the
production rate compared to that of a standard-model Higgs which may be enough
to disentangle the signal from the QCD backgrounds.

\section*{Acknowledgments}

\indent\indent We are grateful for correspondence with D.~Froidevaux,
J.~Incandela, S.~Kuhlmann, E.~Ros, and A.~Rozanov regarding charm mistagging.
This work was supported in part by the U.~S.~Department of Energy under
contract No.~DOE~DE-FG02-91ER40677. We gratefully acknowledge the support of
GAANN, under Grant No.~DE-P200A980724, from the U.~S.~Department of Education
for K.~P.

\section*{Appendix A}
\label{appA}

\begin{figure}[!t]
\begin{center}
\vspace*{0cm} \hspace*{0cm} \epsfxsize=16.0cm
\epsfbox{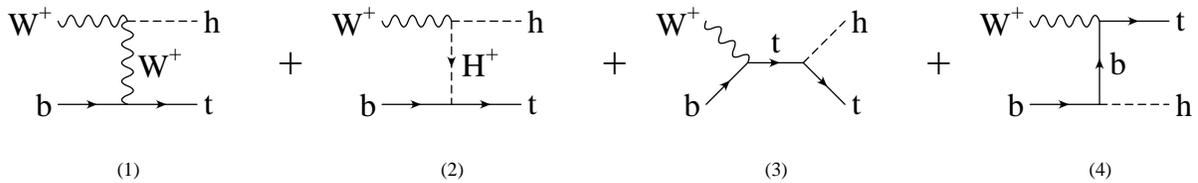} \vspace*{0cm} \caption{Diagrams
contributing to $W^+\, b \to h \,t $ in the 2HDM.}
\label{fig:t-chan-2HDM}
\end{center}
\end{figure}

In this appendix we consider the case of $t$-channel production in
a generic two-Higgs-doublet model (2HDM).
Using the effective-$W$ approximation, we show that
the amplitudes represented by the diagrams in Fig.~\ref{fig:t-chan-2HDM}
contain terms that grow with energy. Nevertheless, the
unitarity of the model implies that these terms must cancel in the
final result, as we show explicitly. In a generic 2HDM that
is invariant under $SU(2)_L \times U(1)_Y$ and conserves CP,
the scalar fields $\Phi_{1,2}$ are doublets of $SU(2)_L$ with hypercharge $Y=1$ and
they develop vacuum expectation values $v_{1,2}$ that break $SU(2)_L \times U(1)_Y$ to
$U(1)_{\rm EM}$. This results in a mass $m_W^2=\frac{1}{4} g^2 v^2$ and
$m_Z^2=\frac{1}{4} (g^2+g'^2) v^2$ with $v^2=v_1^2+v_2^2=(\sqrt{2} G_F)^{-1}$.
The particle content can be exploited to fully parameterize the model.
In addition to $\tan \beta=v_2/v_1$,
we can use the masses of the four scalars $h,H,A,H^\pm$,
the mixing angle $\alpha$ between the CP-even states $h,H$,  and
one of couplings appearing in the quartic potential.
The inclusion of the fermions must be done with care in order to suppress tree-level
flavor-changing neutral currents. One common choice is to impose a
discrete symmetry in such a way that $\Phi_1$ couples only to down-type quarks
and leptons while $\Phi_2$ couples only to up-type quarks~\cite{Glashow:1977nt}. This way
of coupling the Higgs fields to the fermions is the same as in the
minimal supersymmetric standard model and is called type II.

The contributions from the four diagrams in Fig.~\ref{fig:t-chan-2HDM}
read
\begin{eqnarray}
i {\cal A}_1 &=& i\; \frac{g \, g_{WWh}\, m_W}{2\sqrt{2}} \, \;
\bar{u}(p_t)\,  \gamma^{\mu}(1-\gamma^5)\, u(p_b) \cdot
\frac{g_{\mu\nu}-\frac{(p_b-p_t)_\mu (p_b-p_t)_\nu}{m_W^2}}
{(p_b-p_t)^2-m_W^2}\cdot \epsilon_W^\nu\,,\\
i {\cal A}_2 &=&  i\; \frac{g \,g_{WH^+h}}{2\sqrt{2}}\;
\bar{u}(p_t)\; \left[
\frac{m_t}{m_W}\, {\rm cot }\beta \,(1-\gamma^5)+
\frac{m_b}{m_W}\, {\rm tan }\beta \,(1+\gamma^5) \right] \;
u(p_b)\times \nonumber\\
&& \frac{\epsilon_W \cdot (p_t-p_b-p_h)}{(p_b-p_t)^2-m_{H^+}^2}\,,\\
i {\cal A}_3 &=& \!\!-i \; \frac{g\,g_{tth} }{2\sqrt{2}} \;
\frac{\bar{u}(p_t)(\slash \!\!\! p_t+ \slash \!\!\! p_h +m_t)
\slash \!\!\!\epsilon_W (1-\gamma^5) u(p_b)}{(p_t+p_h)^2-m_t^2}\,,\\
i {\cal A}_4 &=& \!\!-i \; \frac{g \, g_{bbh}}{2\sqrt{2}} \;
\frac{\bar{u}(p_t) \slash \!\!\!\epsilon_W
(1-\gamma^5) (\slash \!\!\! p_b- \slash \!\!\! p_h +m_b)
u(p_b)}{(p_b-p_h)^2-m_b^2}\,,
\end{eqnarray}
which in the high-energy limit ($s,-t,-u\gg m_h^2,m_{H^+}^2,m_W^2,m_t^2$)
and for a longitudinally-polarized $W$ ($\epsilon_W^\mu \simeq p_W^\mu/m_W $)
reduce to
\begin{eqnarray}
i {\cal A}_1 &\sim& i \frac{ g\, g_{WWh}}{4 \sqrt{2}\, m_W^2} \, \;
 \bar{u}(p_t) \; \left[
        m_b \, (1+\gamma^5)\,
      - m_t \, (1-\gamma^5)\,
        \right] \; u(p_b)\,,\\
i {\cal A}_2 &\sim& i \frac{ g \, g_{WH^+h}}{2 \sqrt{2}\, m_W^2}\, \;
\bar{u}(p_t) \; \left[
        m_b \,{\rm tan }\beta \, (1+\gamma^5)\,
      + m_t \,{\rm cot }\beta \, (1-\gamma^5)\,
                \right] \; u(p_b)\,,\\
i {\cal A}_3 &\sim&\!\!\!-i \frac{ g \, g_{tth}}{2 \sqrt{2}\, m_W}\, \;
\bar{u}(p_t) \; (1-\gamma^5)\,  \; u(p_b)\,,\\
i {\cal A}_4 &\sim&i \frac{ g \, g_{bbh}}{2 \sqrt{2}\, m_W}\, \;
\bar{u}(p_t) \; (1+\gamma^5)\,  \; u(p_b)\,.
\end{eqnarray}
Unitarity therefore requires that the following relations hold true:
\begin{eqnarray}
&&~~\frac{g_{WWh}}{2}\, m_b + g_{WH^+h}\, {\rm tan }\beta\; m_b + g_{bbh}\; m_W= 0 \,,\\
&&- \frac{g_{WWh}}{2}\, m_t + g_{WH^+h}\, {\rm cot }\beta\; m_t - g_{tth}\; m_W= 0 \,.
\end{eqnarray}
That this is indeed the case can be easily verified using the couplings of the 2HDM,
\begin{eqnarray}
g_{WWh} &=& ~~ g ~\sin( \beta-\alpha) \,,\\
g_{WH^+h} &=& -\frac{g}{2} \cos( \beta-\alpha) \,,\\
g_{tth} &=& -  \frac{g m_t}{2 m_W}\; \frac{\cos\alpha}{\sin\beta}\,,\\
g_{bbh} &=& ~~ \frac{g m_b}{2 m_W}\; \frac{\sin\alpha}{\cos\beta}\,.
\end{eqnarray}
%%%%%%%%%%%%%%%%%%%%%%%%%%%%%%%%%%%%%%%%%%%%%%%%%%%%%%%%%%%%%%%%%%%%%%%%%%%%%
%%%%%%%%%%%%%%%%%%%%%%%%%%%%%%%%%%%%%%%%%%%%%%%%%%%%%%%%%%%%%%%%%%%%%%%%%%%%%
Analogous relations can be derived for the production of the heavy
neutral Higgs $H$ and the results can be obtained from those above
with the replacement $\alpha \to \alpha- \frac{\pi}{2}$. The
production of the CP-odd state $A$ differs from that of the
CP-even Higgs bosons in that its coupling to the $W$ boson is
zero. In this case the divergent terms coming from the diagrams
where the Higgs couples to the quarks cancel with those coming
from the second diagram in Fig.~\ref{fig:t-chan-2HDM}. An explicit
calculation gives:
\begin{eqnarray}
i {\cal A}_2 &\sim&  \frac{ g \, g_{WH^+A}}{2 \sqrt{2}\, m_W^2}\, \;
\bar{u}(p_t) \; \left[
        m_b \,{\rm tan }\beta \, (1+\gamma^5)\,
      + m_t \,{\rm cot }\beta \, (1-\gamma^5)\,
                \right] \; u(p_b)\,,\\
i {\cal A}_3 &\sim&\!\!\!-\frac{ g \, g_{ttA}}{2 \sqrt{2}\, m_W}\, \;
\bar{u}(p_t) \; (1-\gamma^5)\,  \; u(p_b)\,,\\
i {\cal A}_4 &\sim&\!\!\!- \frac{ g \, g_{bbA}}{2 \sqrt{2}\, m_W}\, \;
\bar{u}(p_t) \; (1+\gamma^5)\,  \; u(p_b)\,.
\end{eqnarray}
Unitarity entails that
\begin{eqnarray}
&& g_{WH^+A}\, {\rm tan }\beta\; m_b + g_{bbh}\; m_W= 0 \,,\\
&& g_{WH^+A}\, {\rm cot }\beta\; m_t + g_{tth}\; m_W= 0 \,.
\end{eqnarray}
The above constraints are satisfied by the couplings of the 2HDM,
\begin{eqnarray}
g_{WH^+A} &=& \frac{g}{2} \,,\\
g_{ttA} &=&\!\!\! -  \frac{g m_t}{2 m_W}\; \cot \beta\,,\\
g_{bbA} &=&\!\!\! -  \frac{g m_b}{2 m_W}\; \tan \beta\,.
\end{eqnarray}

\end{document}